\documentstyle[aps,preprint,psfrag,epsfig]{revtex}

\begin{document}
\draft

\bigskip

\title{ Phases of the Brans-Dicke Cosmology with Matter}
\author{ Chanyong Park and Sang-Jin Sin }
\address{Department of Physics, Hanyang University \\ Seoul, Korea}
\maketitle
\tighten
\begin{abstract}

We study the cosmology  of the  Brans-Dicke
theory with perfect fluid type matter.
In our previous work, we found exact solutions
for any Brans-Dicke parameter $\omega$ and for general parameter $\gamma$ of
equation of state.
In this paper we further
study the cosmology of these solutions by analyzing them
according to their asymptotic behaviors.
The cosmology is classified into 19 phases
according to the values of $\gamma$ and $\omega$.
The effect of the cosmological constant to the Brans-Dicke theory
is a particular case of our model.
We give plot of time evolution of the scale factor by numerical investigations.
We also give a comparison of the solutions for the theories with and
without matter.

\end{abstract}


\newpage

\section{Introduction}
Recent developments of the string theory suggest that in a
regime  of Planck length curvature, quantum fluctuation is very large
so that string coupling becomes large and consequently
the fundamental string degrees of freedom
are not weakly coupled {\it good} ones \cite{witten}.
Instead, solitonic degrees of freedom like p-brane or
D-p-brane\cite{pol} are more important.
Therefore it is a very interesting question
to ask what is the effect of these new degrees of freedom to
the space time structure especially whether including these degrees
of freedom resolve the initial singularity, which is a problem in
standard general relativity.
The new gravity theory that can deal with such new degree of freedom should be
a deformation of standard general relativity so that in a certain limit it should be
reduced to the standard Einstein theory.  The
Brans-Dicke theory\cite{brans} is a generic deformation of the general relativity
allowing variable gravity coupling. Therefore whatever is the motivation to modify
the Einstein theory, the Brans-Dicke theory is the first one to be considered.
As an example, low energy limit of the string theory
contains the Brans-Dicke theory with a fine tuned deformation parameter
($\omega$=-1).

Without knowing the exact theory of the p-brane cosmology,
the best guess is that it should be a Brans-Dicke theory with matters.
In fact there is a some evidence for this \cite{du}, where it is found that the
natural metric that couples to the p-brane is the  Einstein metric
multiplied by certain power of dilaton field.  In terms of this new
metric, the action that gives the p-brane solution  becomes
Brans-Dicke action with definite deformation  parameter $\omega$
depending on p. Using this action, Rama \cite{ra} recently  argued
that the gas of  solitonic  p-brane \cite{du} treated as
a perfect fluid type matter in a Brans-Dicke theory can resolve the
initial singularity without any explicit solution.
In a previous papers \cite{ps,sgl}, we have studied this model and found
exact cosmological solutions  for any Brans-Dicke parameter $\omega$ and for
general equation of  state and classify the cosmology of the
solutions according to the range of parameters involved.

In this paper we further
study the cosmology of these solutions by analyzing them
according to their asymptotic behaviors.
The cosmology is classified into 19 phases
according to the values of $\gamma$ and $\omega$.
The effect of the cosmological constant to the Brans-Dicke theory
is a particular case of our model.
We give plot of time evolution of the scale factor by numerical investigations.
We also give a comparison of the solutions for the theories with and
without matter.

The rest of this paper is organized as follows.
In section II,  we set up the notation and review the result of our  previous results
\cite{ps,sgl}.
In section III, we describe two new phases which were not mentioned in
\cite{ps} and classify the cosmology into $19$ phases. By the numerical work
as well as the analytical method, the
behaviors of the scale factor is presented explicitly by figures.
In section IV, we summarize and conclude with some discussions.

\section{Brans-Dicke cosmology with matter}
First, we briefly review our earlier work\cite{ps}.
We consider the Brans-Dicke theory and
analyze the evolution of the  $D$ dimensional homogeneous isotropic
universe with the perfect fluid type matter.
The action is given by
\begin{equation}
S = \int d^D x \sqrt{-g} e^{-\phi} \left[ {\cal R} - \omega
\partial_{\mu} \phi \partial^{\mu} \phi \right] + S_m ,
\end{equation}
where $\phi$ is the dilaton field and $S_m$ is the
matter part of the action. Here we assume that the matter has no
dilaton coupling.

Let's assume that the matter can be treated as a perfect fluid
with the equation of state
\begin{equation}
p=\gamma \rho, \gamma <1.  \label{eos}
\end{equation}
Therefore our starting point is the equation of the Brans-Dicke
theory\cite{Weinberg,Veneziano}
\begin{eqnarray}
{\cal R}_{\mu\nu} - \frac{g_{\mu\nu}}{2} {\cal R} &=&
       \frac{e^{\phi}}{2} T_{\mu\nu} + \omega  \{ \partial_{\mu} \phi
       \partial_{\nu} \phi - \frac{g_{\mu\nu}}{2} (\partial \phi)^2
       \} \nonumber \\ && + \{ -\partial_{\mu} \partial_{\nu} \phi +
       \partial_{\mu} \phi \partial_{\nu} \phi + g_{\mu\nu} {\cal D}^2
       \phi - g_{\mu\nu} (\partial \phi)^2 \} , \nonumber \\
 0 &=&{\cal R} - 2\omega {\cal D}^2 \phi + \omega (\partial \phi)^2
\end{eqnarray}
where $\phi$ is the dilaton and ${\cal D}$ means a covariant
derivative. ${\cal R}$ is the curvature scalar and the metric
is given as the follows:
\begin{equation}
ds^2 = -\frac{1}{\cal N} dt^2 + e^{2\alpha(t)} \delta_{ij} dx^i dx^j \;\;
( i,j = 1, 2, \cdots, D-1) ,
\end{equation}
where $e^{\alpha(t)} (=a(t))$ is the scale factor and ${\cal N}$ is
the (constant) lapse  function.

The energy-momentum tensor is given by
\begin{equation}
T_{\mu\nu} = p g_{\mu\nu} + (p + \rho) U_{\mu} U_{\nu}
\end{equation}
where $U_{\mu}$ is the fluid velocity. The hydrostatic equilibrium
condition of energy-momentum conservation is
\begin{equation}
\dot{\rho} + (D-1) (p + \rho) \dot{\alpha} = 0 . \label{sol1}
\end{equation}
Using  the equation of state, eq.(\ref{eos}), with a
free parameter $\gamma$, we get the solution
\begin{equation}
\rho = \rho_0 e^{-(D-1)(1+\gamma) \alpha},
\end{equation}
where $\rho_0$ is a real number.
Our goal is to study how the
metric variables change their behavior for various values of
$\gamma$ and $\omega$.
Now, since we consider only the time dependence, the action can be brought to
the following form
\begin{eqnarray}
S &=& \int dt \;\; e^{(D-1)\alpha-\phi} \left[ \frac{1}{\sqrt{\cal N}} \big{\{}
      -(D-2)(D-1)\dot{\alpha}^2 + 2 (D-1)\dot{\alpha}\dot{\phi} + \omega\dot{\phi}^2
      \big{\}} \right. \nonumber \\
  && \left. \hspace{2.5cm} - {\sqrt{\cal N}} \rho_0 e^{-(D-1)(1+\gamma)\alpha + \phi}
     \frac{}{} \right].
\end{eqnarray}
where we eliminated $p$ and $\rho$ by eq.(\ref{sol1}).
After getting the constraint equation by varying over the constant lapse function, ${\cal N}$, we can set it to be 1.

Now, introducing a new time variable $\tau$ by
\begin{equation}
dt = e^{(D-1)\alpha - \phi} d\tau,
\end{equation}
and  the new variables
\begin{eqnarray}
 X    &=& -\frac{1}{2}[(D-1)(1-\gamma) \alpha - \phi] , \nonumber \\
 Y      &=& \alpha + \frac{\nu}{\kappa} X,
\end{eqnarray}
 the action can be written as
\begin{equation}
S= \int d\tau \left[ \frac{1}{\sqrt{\cal N}} \big{\{} (D-1) \kappa \dot{Y}^2 + \mu
     \dot{X}^2 \big{\}} - \sqrt{\cal N} \rho_0 e^{-2X} \right] ,
\end{equation}
where
\begin{eqnarray}
 \kappa &=& (D-1)(1-\gamma)^2 (\omega-\omega_{\kappa}) , \nonumber \\
 \nu    &=& 2(1-\gamma)(\omega-\omega_{\nu}) , \nonumber \\
 \mu    &=& - \frac{4(D-2)}{\kappa} (\omega-\omega_{-1}) , \nonumber \\
 \omega_{\kappa} &=& -\frac{D-2D\gamma+2\gamma}{(D-1)(1-\gamma)^2}, \nonumber \\
 \omega_{\nu}    &=& -\frac{1}{1-\gamma}, \nonumber \\
 \omega_{-1}     &=& -\frac{D-1}{D-2}.
\end{eqnarray}
The constraint equation is given by
\begin{equation}
0 = (D-1)\kappa \dot{Y}^2 + \mu \dot{X}^2 + \rho_0 e^{-2X}.
\end{equation}
Note that for $D>2$ and $\gamma <1$, the sign of $\kappa$ is
determined by that of $\omega -\omega_{\kappa}$ and
the  sign of $\mu$ is
determined by that of $\omega -\omega_{-1}$ and $\kappa$.

 The equations of
motion are simply
\begin{eqnarray}
0 &=& \ddot{Y} , \nonumber \\
0 &=& \ddot{X} - \frac{\rho_0}{\mu} e^{-2X} .
\end{eqnarray}
When $\rho_0 = 0$, the situation is that of the string cosmology
discussed first in \cite{vene} and
the solution for $X$ is $X = c \tau$.
One can easily show that this solution
has two disconnected branches in terms of the original time $t$;
one is inflation-type and the other is FRW-type.
If $\rho_0 \neq 0$, the asymptotic behavior of $X$ is
$X \sim c \mid \tau \mid$ as we will see later. In other word,
the behavior of cosmology at $\rho_0 \neq 0$ is not
continuously connected to that of cosmology at $\rho_0 =0$
in $\tau \to -\infty$ region. Some of the new aspects of the
cosmology due to the presence of the matter come from this
discontinuity. If $\rho_0$ is a negative constant, then the
solution oscillates in time, leading to unphysical solution.
This includes the situation where there is a negative cosmological constant
in the Brans-Dicke theory.
In this paper, therefore, we will consider only positive $\rho_0$.

If $\omega$ is less than $\omega_{-1}$, the kinetic term
of the dilaton has a negative energy in Einstein frame. So we will
consider the case where $\omega$ is larger than $\omega_{-1}$.
According to the sign of $\kappa$, the types of solutions are very
different. When $\kappa$ is negative, the exact solution is
\begin{eqnarray}
X &=& \ln \big[ \frac{q}{c} \cosh(c\tau) \big], \nonumber \\
Y &=& A \tau + B ,
\end{eqnarray}
where $c$, $A$, $B$ and $q = \sqrt{\frac{\rho_0}{\mid \mu \mid}}$ are
arbitrary real constants. Using the constraint equation, we can determine
$A$ in terms of other parameters,
\begin{equation}
A = \frac{c}{\delta}, \hspace{1cm} \rm{with} \hspace{0.5cm}
\delta=\sqrt{ - \frac{(D-1)\kappa}{\mu}} = \frac{\mid \kappa \mid}{2\sqrt{1+
\omega \frac{D-2}{D-1}}} .
\end{equation}
If $\kappa$ is zero, it turns out that the solution of the equations of
motion does not satisfy the constraint equation. If $\kappa$
is positive, the solution is
\begin{eqnarray}
 X &=& \ln \big[ \frac{q}{c} \mid \sinh(c\tau) \mid \big], \nonumber \\
 Y &=& \frac{c}{\delta} \tau + B.
\end{eqnarray}

\section{ Phases of the Brans-Dicke theory}

In \cite{ps}, the behaviors of the scale factor were classified
by $16$ phases according to $\omega$ and $\gamma$, see figure
1, and we showed that the asymptotic behaviors of scale factor
$a(\tau)$ and time $t(\tau)$ at $\tau \rightarrow \pm \infty$ are as follows:
\begin{eqnarray}
t-t_0 &\approx& \frac{1}{T_{\pm}} \left( e^{T_{\pm} \tau} - e^{T_{\pm}
              \tau_0} \right) \nonumber \\
T_{\pm} &=& \frac{2c}{\mid \kappa \mid} \left[ (D-1)\gamma \sqrt{1+\omega
            \frac{D-2}{D-1}} \mp {\rm sign}(\kappa) \{\kappa+(D-1) \gamma (
            1 + \omega(1-\gamma))\} \frac{}{} \right], \nonumber \\
a(\tau) &\approx& e^{H_{\pm} \tau} \nonumber \\
H_{\pm} &=& \frac{2c}{\mid \kappa \mid} \left[\sqrt{1+\omega\frac{D-2}{D-1}}
            \mp {\rm sign}(\kappa) \{ 1 + \omega (1-\gamma) \}
              \right]. \label{asym}
\end{eqnarray}
Note that the range of $t$ is determined by the sign of $T_{\pm}$:
\begin{eqnarray*}
(-\infty, \infty) \;\; & \rm{if} \;\; & T_- < 0 < T_+ , \\
(-\infty, t_f ) \;\; & \rm{if} \;\; & T_- < 0 \; \rm{and} \; T_+ <0 , \\
(t_i , \infty) \;\; & \rm{if} \;\; & T_- >0 \; \rm{and} \; T_+ >0 ,
\\ (t_i , t_f )   \;\; & \rm{if} \;\; & T_+ < 0 < T_- .
\end{eqnarray*}
For $\kappa >0$, $t(\tau)$ and
$a(\tau)$ behave as \cite{ps}
\begin{eqnarray}
t &\approx& - {\rm sign}(\tau) \frac{q^{-\eta} e^{(D-1) \gamma B}}{(\eta-1)}
            \frac{1}{\mid \tau \mid^{\eta-1}} , \nonumber \\
a &\approx&  e^B (q  \mid \tau \mid)^{-\frac{2(1-\gamma)(\omega
            -\omega_{\nu})}{\mid \kappa  \mid}},
\end{eqnarray}
as $\tau$ goes to zero,
where $\eta = 2+\frac{(D-1)\gamma \nu}{\kappa}$. $t(\tau)$ is singular
at $\tau \rightarrow 0$ if $\eta >1$.
So for $\kappa >0$ and $\eta >1$, the scale factor $a(t)$ has
two branches. The asymptotic form of $a(t)$ as a function of $t$ is given by
\begin{eqnarray}
a(t) \approx [T_- (t- t_i )]^{H_- / T_-} \;\; &\rm{at}& \;\; \tau
\rightarrow -\infty,  \nonumber \\
a(t) \approx [T_+ (t- t_f )]^{H_+ / T_+} \;\; &\rm{at}& \;\; \tau
\rightarrow \infty,
\end{eqnarray}
where $t_i$ ($t_f$) is a starting (ending) point at a finite
time. Eq.(\ref{asym}) contains the cases where $t$ starts from $-\infty$
and/or ends at $\infty$ by setting $t_i =0$ and/or
$t_f =0$. According to the sign of $T_{\pm}$ and $H_{\pm} /T_{\pm}$ and
the singularity at $\tau=0$, we classified the behavior
of the Brans-Dicke theory\cite{ps}.
Here we summarize the result by table 1. \\

\vspace{0.3cm}
\begin{tabular}{|c||c|c|c|c|c|c|}
\hline
phase  & sign of $\kappa$ & sign of $T_-$ & sign of $T_+$ &
 range of $t$ & sign of $H_- /T_-$ & sign of $H_+ /T_+$ \\
\hline \hline
 $I$         & - & + & - & $[t_i,t_f]$    & + & +   \\
\hline
 $II$        & - & - & - & $(-\infty,t_f]$    & - & +   \\
\hline
 $III^-$     & + & + &   & $[t_i,t_f]$    & + &     \\
\hline
 $III^+$     & + &   & + & $[t_i,\infty)$     &   & +   \\
\hline
 $IV$        & - & - & + & $(-\infty,\infty)$ & + & +   \\
\hline
 $V$         & - & + & + & $[t_i,\infty)$     & - & +   \\
\hline
 $VI$        & - & + & + & $[t_i,\infty)$     & + & +   \\
\hline
 $VII^-$     & + & + &   & $[t_i,\infty)$     & + &     \\
\hline
 $VII^+$     & + &   & + & $(-\infty,\infty)$ &   & +   \\
\hline
 $VIII^-$    & + & + &   & $[t_i,\infty)$     & + &     \\
\hline
 $VIII^+$    & + &   & - & $(-\infty,t_f]$    &   & -   \\
\hline
 $IX^-$      & + & + &   & $[t_i,\infty)$     & + &     \\
\hline
 $IX^+$      & + &   & + & $(-\infty,\infty)$ &   & +   \\
\hline
 $X^-$       & + & + &   & $[t_i,\infty)$     & + &     \\
\hline
 $X^+$       & + &   & - & $(-\infty,t_f]$    &   & -   \\
\hline
 $XI^-$      & + & + &   & $[t_i,\infty)$     & + &     \\
\hline
 $XI^+$      & + &   & - & $(-\infty,t_f]$    &   & +   \\
\hline
\end{tabular}

\vspace{0.3cm} Table 1. The sign of $T_{\pm}$ determines the range of
time $t$ as following;  $t(\tau)$ maps the real line of $\tau$ to
(1) if $T_- < 0 < T_+$, $(-\infty, \infty)$ (2) if $T_- <0$ and $T_+
<0$, $(-\infty, t_f]$ (3) if $T_- >0$ and $T_+ >0$, $[t_i , \infty)$
(4) if $T_+ < 0 < T_-$, $[t_i , t_f ]$. The sign of $H_{\pm}
/T_{\pm}$ determines the asymptotic behavior of scale factor $a(t)$.

\vspace{0.5cm}
Now, notice that not only the sign of $H_{\pm} /T_{\pm}$ but also
that of $H_{\pm} /T_{\pm}-1$ is important because the
 the universe will accelerate if $H_{\pm} /T_{\pm} -1 >0$ and
decelerate if $H_{\pm} /T_{\pm} -1 <0$ when $\tau \rightarrow \pm 
\infty$.
Therefore, we further classify the phases of cosmology accordingly.

\subsection{ Case $\omega < \omega_{\kappa}$}

\subsubsection{ $H_- /T_- > 1$}
\begin{itemize}
  \item  For $T_- >0$, the condition $H_- /T_- > 1$ is reduced to
\begin{equation}
\sqrt{1+\omega \frac{D-2}{D-1}} > \frac{(D-1)\gamma
-1}{(1-\gamma)(D-1)}. \label{ineq}
\end{equation}
If $\gamma < 1/(D-1)$, the
condition is automatically satisfied. If $\gamma > 1/(D-1)$,
inequality (\ref{ineq}) turns out to be reduced to $\omega >
\omega_{\kappa}$, which is surprising. This means that there is no
solution. Therefore among the regions $I$ and $VI$ which have $T_-
>0$, $H_- >0$ and $\omega < \omega_{\kappa}$, only $I$ satisfies
$H_- /T_- > 1$.
  \item
  If $T_- <0$, the condition $H_- /T_- > 1$ is reduced to
\begin{equation}
\sqrt{1+\omega \frac{D-2}{D-1}} < \frac{(D-1)\gamma
-1}{(1-\gamma)(D-1)},
\end{equation}
whose solution is $\gamma > 1/(D-1)$ and $\omega <\omega_{\kappa}$.
Only $IV$ satisfies conditions, $\omega < \omega_{\kappa}$,
$T_- <0$, $H_- <0$ and $H_- /T_- >1$.
\end{itemize}

\subsubsection{ $H_+ /T_+ > 1$}

\begin{itemize}
  \item
   If $T_+ >0$, the condition $H_+ /T_+ > 1$ implies
\begin{equation}
\sqrt{1+\omega \frac{D-2}{D-1}} < -\frac{(D-1)\gamma
-1}{(1-\gamma)(D-1)},
\end{equation}
whose solution is given by $\gamma < 1/(D-1)$ and $\omega
<\omega_{\kappa}$. There is no region satisfying
$T_+ >0$, $\gamma < 1/(D-1)$ and $\omega < \omega_{\kappa}$.
  \item
 If $T_+ <0$, $H_+ /T_+ >1$ is reduced to
\begin{equation}
\sqrt{1+\omega \frac{D-2}{D-1}} > -\frac{(D-1)\gamma
-1}{(1-\gamma)(D-1)},
\end{equation}
whose solution is $\gamma > 1/(D-1)$ or $\omega >\omega_{\kappa}$
for $\gamma < 1/(D-1)$.
There is no region satisfying the
conditions, $\omega <\omega_{\kappa}$, $T_+ <0$ and $H_+ /T_+ >1$.
\end{itemize}

\subsection{ Case $\omega > \omega_{\kappa}$}

\subsubsection{ $H_- /T_- >1$}

\begin{itemize}
  \item  For $T_- >0$, the condition $H_- /T_- >1$ is
reduced to
\begin{equation}
\sqrt{1+\omega \frac{D-2}{D-1}} < -\frac{(D-1)\gamma
-1}{(1-\gamma)(D-1)},
\end{equation}
whose solution is $\gamma < 1/(D-1)$ and $\omega < \omega_{\kappa}$.
Therefore there is no solution satisfying the conditions,
$\omega < \omega_{\kappa}$ and $H_- /T_- >1$.
  \item
If $T_- < 0$, the condition is given by
\begin{equation}
\sqrt{1+\omega \frac{D-2}{D-1}} > -\frac{(D-1)\gamma
-1}{(1-\gamma)(D-1)}.
\end{equation}
The solution is $\gamma > 1/(D-1)$ or $\omega >\omega_{\kappa}$ for
$\gamma < 1/(D-1)$. Therefore, the solution
is summarized by $\omega >\omega_{\kappa}$.
But in the case $\omega >\omega_{\kappa}$,
there is no region satisfying $T_- <0$.
\end{itemize}

\subsubsection{  $H_+ /T_+ >1$}

\begin{itemize}
  \item  For $T_+ >0$, the condition $H_+ /T_+ >1$ is
reduced to
\begin{equation}
\sqrt{1+\omega \frac{D-2}{D-1}} > \frac{(D-1)\gamma
-1}{(1-\gamma)(D-1)}.
\end{equation}
The solution is $\gamma < 1/(D-1)$ or $\omega > \omega_{\kappa}$ for
$\gamma > 1/(D-1)$. So the solution is summarized by $\omega
>\omega_{\kappa}$ like the last case.
The regions $III^+$, $VII^+$, $IX^+$ have the
solution satisfying $\omega >\omega_{\kappa}$, $T_+ >0$ and
$H_+ /T_+ >1$.
  \item
 If $T_+ <0$, the above condition is reduced to
\begin{equation}
\sqrt{1+\omega \frac{D-2}{D-1}} < \frac{(D-1)\gamma
-1}{(1-\gamma)(D-1)},
\end{equation}
whose solution is $\gamma > 1/(D-1)$ and $\omega <\omega_{\kappa}$.
Therefore, there is no solution satisfying $\omega <\omega_{\kappa}$,
because the solution $\omega <\omega_{\kappa}$ is inconsistent with
the assumption $\omega >\omega_{\kappa}$.
\end{itemize}

\subsubsection{ The power behavior of scale factor at $\tau
\rightarrow 0$}

At $\tau \rightarrow 0$, $a(t)$ is given by
\begin{equation}
a(t) \approx E \times \mid t \mid^{\Gamma} \label{tau0}
\end{equation}
where $\Gamma$ is given by
\[
\Gamma = \frac{2(1-\gamma)(\omega-\omega_{\nu})}{(\eta-1) \kappa}
\]
and a constant $E$ becomes
\[
E = [q(\eta -1)]^{\frac{2(1-\gamma)(\omega-\omega_{\nu})}{(\eta-1)
\mid \kappa \mid}}  e^{B \left[1-  \frac{2(D-1)\gamma (1-\gamma)
(\omega-\omega_{\nu})}{(\eta-1) \kappa} \right] }.
\]
For $\omega >\omega_{\kappa}$ and $\eta >1$, the condition that
$\Gamma$ is positive is satisfied in the region
$\omega > \omega_{\nu}$.
The condition $\Gamma >1$ is reduced to
\begin{equation}
(1-\gamma) + [(D-1) \gamma + (D-3)] \omega + (D-2) < 0.
\end{equation}
This gives the following solution
\begin{eqnarray}
\omega < -\frac{D-2}{(1-\gamma)((D-1)\gamma +D-3)} \;\; &{\rm for}&
 \;\; \gamma > -\frac{D-3}{D-1},\label{gamma+} \\
\omega > -\frac{D-2}{(1-\gamma)((D-1)\gamma +D-3)} \;\; &{\rm for}&
 \;\; \gamma < -\frac{D-3}{D-1} \label{gamma-} .
\end{eqnarray}
Note that for $\gamma > -\frac{D-3}{D-1}$, $\omega_{\kappa}$ is always
greater than $-\frac{D-2}{(1-\gamma)((D-1)\gamma +D-3)}$. So in that case,
there is no solution. As a
result, the solution to $\Gamma>1$ is given by eq.(\ref{gamma-}).
This divide the region XI of figure 1 into two region:
for $0<\Gamma <1$ we call this as region XI and for
$\Gamma >1$ we call this as region XII. See figure 2.

Now we summarize all possible phases in Table 2. \\

\vspace{0.3cm}
\noindent
\begin{tabular}{|c||c|c|c|c|c|c|c|c|}
\hline
phase & sign of & sign of & sign of & \hspace{0.5cm} range
of  \hspace{0.5cm} &  $H_- /T_-$ & \hspace{1cm} $\Gamma$
\hspace{1cm} & $H_+ /T_+$ \\
      & $\kappa$ & $T_-$ & $T_+$ & $t$ &  &($\tau \to 0$) & \\
\hline \hline
$I$     & - & + & - & $[t_i,t_f]$ &
$H_- / T_- >1$ & & $0< H_+ /T_+ <1$ \\
\hline
$II$    & - & - & - &
$(-\infty,t_f]$ & $H_- / T_- <0$ &  & $0< H_+ /T_+ <1$ \\
\hline
$III^-$ & + & + &  & $[t_i,t_f]$ & $0< H_- / T_- <1$ &  & \\
\hline
$III^+$ & + & & + & $[t_i,\infty)$ & & & $H_+ /T_+ >1$ \\
\hline
$IV$    & - & - & + & $(-\infty,\infty)$ & $H_- / T_- >1$ & &
$0< H_+ /T_+ <1$ \\
\hline
$V$     & - & + & + & $[t_i,\infty)$ & $H_-
/ T_- <0$ & & $0< H_+ /T_+ <1$ \\
\hline
$VI$    & - & + & + &
$[t_i,\infty)$ & $0< H_- / T_- <1$ & & $0< H_+ /T_+ <1$ \\
\hline
$VII^-$ & + & + & & $[t_i,\infty)$ & $0< H_- / T_- <1$ &
$\Gamma <0$ & \\
\hline
$VII^+$ & + & & + & $(-\infty,\infty)$ &
& $\Gamma <0$ & $H_+ /T_+ >1$ \\
\hline
$VIII^-$ & + & + & &
$[t_i,\infty)$ & $0< H_- / T_- <1$ & $\Gamma <0$ & \\
\hline
$VIII^+$ & + & & - & $(-\infty,t_f]$ & & $\Gamma <0$ & $H_+ /T_+ <0$
\\
\hline
$IX^-$ & + & + & & $[t_i,\infty)$ & $0< H_- / T_- <1$
& $0< \Gamma <1$ & \\
\hline
$IX^+$ & + & & + &
$(-\infty,\infty)$ & & $0< \Gamma <1$ & $H_+ /T_+ >1$ \\
\hline
$X^-$ & + & + & & $[t_i,\infty)$ & $0< H_- / T_- <1$ & $0< \Gamma
<1$ & \\
\hline
 $X^+$ & + & & - & $(-\infty,t_f]$ & & $0< \Gamma
<1$ & $H_+ /T_+ <0$ \\
\hline
 $XI^-$ & + & + & & $[t_i,\infty)$ &
$0< H_- / T_- <1$ & $0< \Gamma <1$ & \\
\hline
 $XI^+$ & + & & -
& $(-\infty,t_f]$ & & $0< \Gamma <1$ & $0< H_+ /T_+ <1$ \\
\hline
 $XII^-$ & + & + & & $[t_i,\infty)$ & $0< H_- / T_- <1$ &
$\Gamma >1$ & \\
\hline
 $XII^+$ & + & & - & $(-\infty,t_f]$ & &
$\Gamma >1$ & $0< H_+ /T_+ <1$ \\
\hline
\end{tabular}

\vspace{0.3cm} Table 2. All possible phases are classified. Here
phases $XII^-$ and $XII^+$ are new phases.

\vspace{0.5cm}
By the numerical work, the explicit scale factor behaviors of all phases
are shown in following figures. Here we set $D=4$. \\

All phases in figure 3 have no singularity at $\tau =0$ and the phases $II$
and $IV$ have no initial singularity. The asymptotic behavior of scale
factor is determined by $H_{\pm} /T_{\pm}$.
Note that phases $III^-$ and $III^+$, which are continuously connected
at $\tau =0$, are not distinguished in \cite{ps}. Since scale factor
of phases $III^{\pm}$ vanishes at $\tau = 0$, they are divided into
two phases in this paper.
 \\

Each region included in figure 4 and figure 5 has two
branches and each branch defines a different phase.
For example, the earlier asymptotic behavior of $VII^-$ phase (figure
4a) and the
later asymptotic behavior of $VII^+$ phase (figure 4b) are determined by
$H_{\pm} /T_{\pm}$, the later behavior of $VII^-$ phase and
the earlier behavior of $VII^+$ phase are determined by the behavior of scale
factor near $\tau \to \pm 0$.
The phases, which have no initial singularity, are $+$
phases in figure 4.

In figure 5, phases $XI^{\pm}$ and $XII^{\pm}$ are divided by $\Gamma$
which is given by eq.(\ref{tau0}) at $\tau =0$; $0<\Gamma<1$ for
phases $XI^{\pm}$ and $\Gamma >1$ for phases $XII^{\pm}$.

\section{ Discussion and conclusion}

In this paper we studied the effect of the matter in the Brans-Dicke cosmology.
This was motivated\cite{ps} by the string cosmology\cite{vene} with  gas of  solitonic  p-brane
by treating them  as a perfect fluid type matter in Brans-Dicke theory.
In Brans-Dicke theory, the matter has no dilaton coupling. From the
string theory point of view, this means that the matter has R-R
charge. So the matter considered here corresponds to D-brane gas in
string theory context.
With this matter, we found exact cosmological solutions for any
Brans-Dicke parameter
$\omega$ and for general constant $\gamma$ and
classified all possible phases of the solutions
according to the parameters involved. There are new two phases
$XII^{\pm}$ different from $XI^{\pm}$ for the behavior of scale factor
at $\tau = 0$. So the number of total phases is $19$ and
some of them have no initial singularity. We studied all the phases of
cosmology numerically and gave the figures for the time evolution of the
scale factor.

Recently, ref. \cite{rey} argued that
Holographic principle in the presence of a cosmological constant
 might imply the absence of initial singularity.
In the Brans-Dicke cosmology, we do find some  solutions
avoiding the initial singularity. 
However, when we regard the Brans-Dicke theory as
a string cosmology, we might ask whether there are solutions which resolve
the initial singularity and the graceful exit problem at the same time.
However, the cosmological constant in Brans-Dicke theory is not a
cosmological constant in string theory where the cosmological constant
couples to the dilaton. To discuss the problem in our
frame work, we have to consider the matter coupled with dilaton.
We will discuss the problem in later publications.\\

\noindent{\bf \Large Acknowledgments}
This work has been supported by Hanyang University, Korea, made in
the program year of 1998.

\newpage

\begin{figure}
  \unitlength 1mm
   \begin{center}
      \begin{picture}(70,120)
      \put(-15,10){\epsfig{file=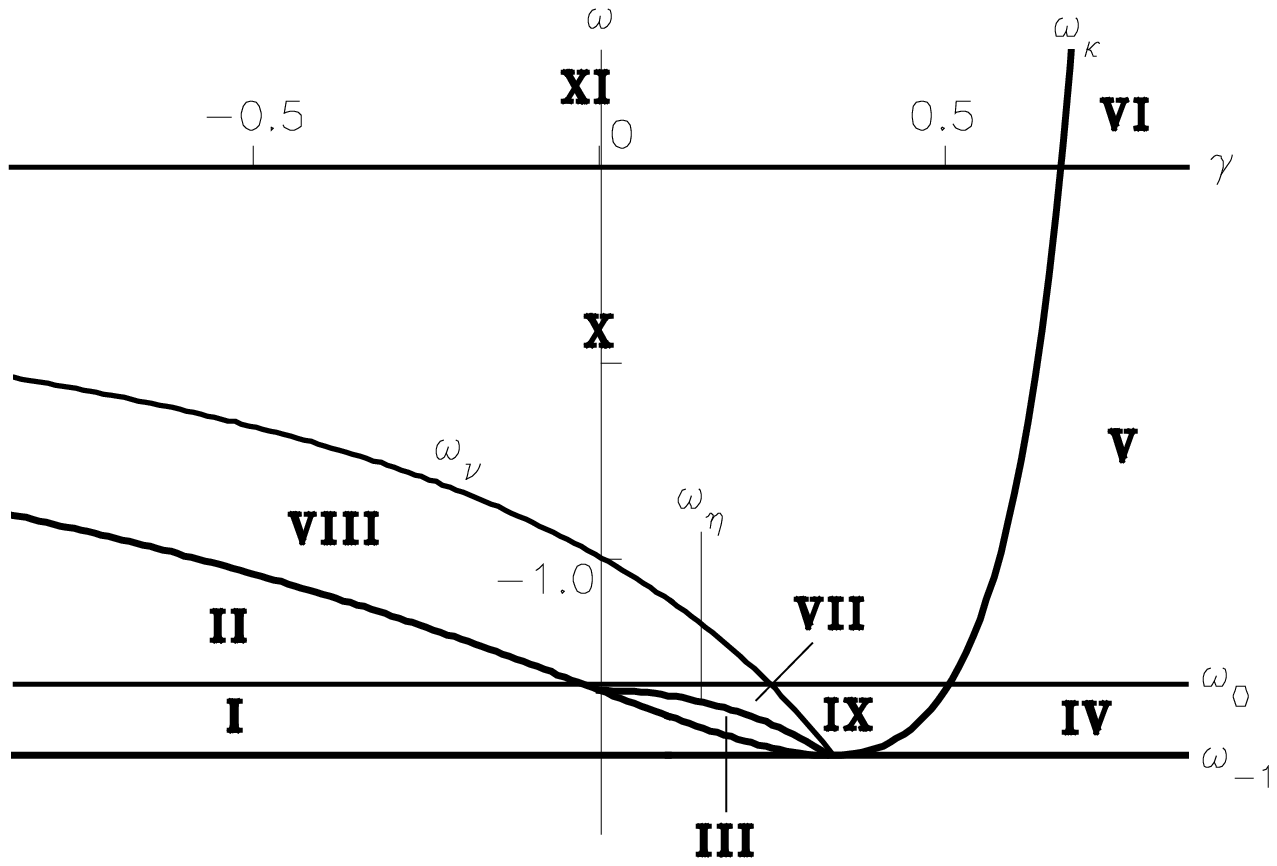,width=10cm,height=8cm}}
      \end{picture}
   \end{center}
\caption{  Phase diagram of 11 regions }
\label{fig1}
\end{figure}

%

\newpage

\vspace{-3cm}
\begin{figure}
  \unitlength 1mm
   \begin{center}
      \begin{picture}(70,130)
      \put(-45,-60){\epsfig{file=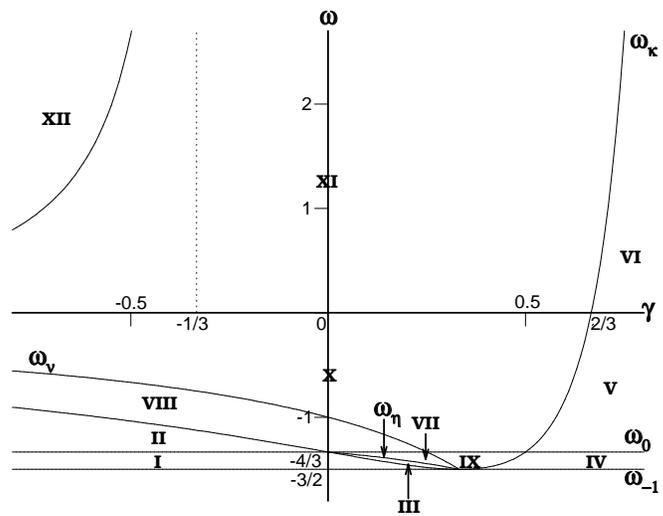,width=14cm,height=20cm}}
      \end{picture}
   \end{center}
\caption{  Phase diagram, 12 regions }
\label{fig1}
\end{figure}

\newpage
\vspace{-3cm}
\begin{figure}
  \unitlength 1mm
   \begin{center}
      \begin{picture}(70,100)
      \put(-50,30){\epsfig{file=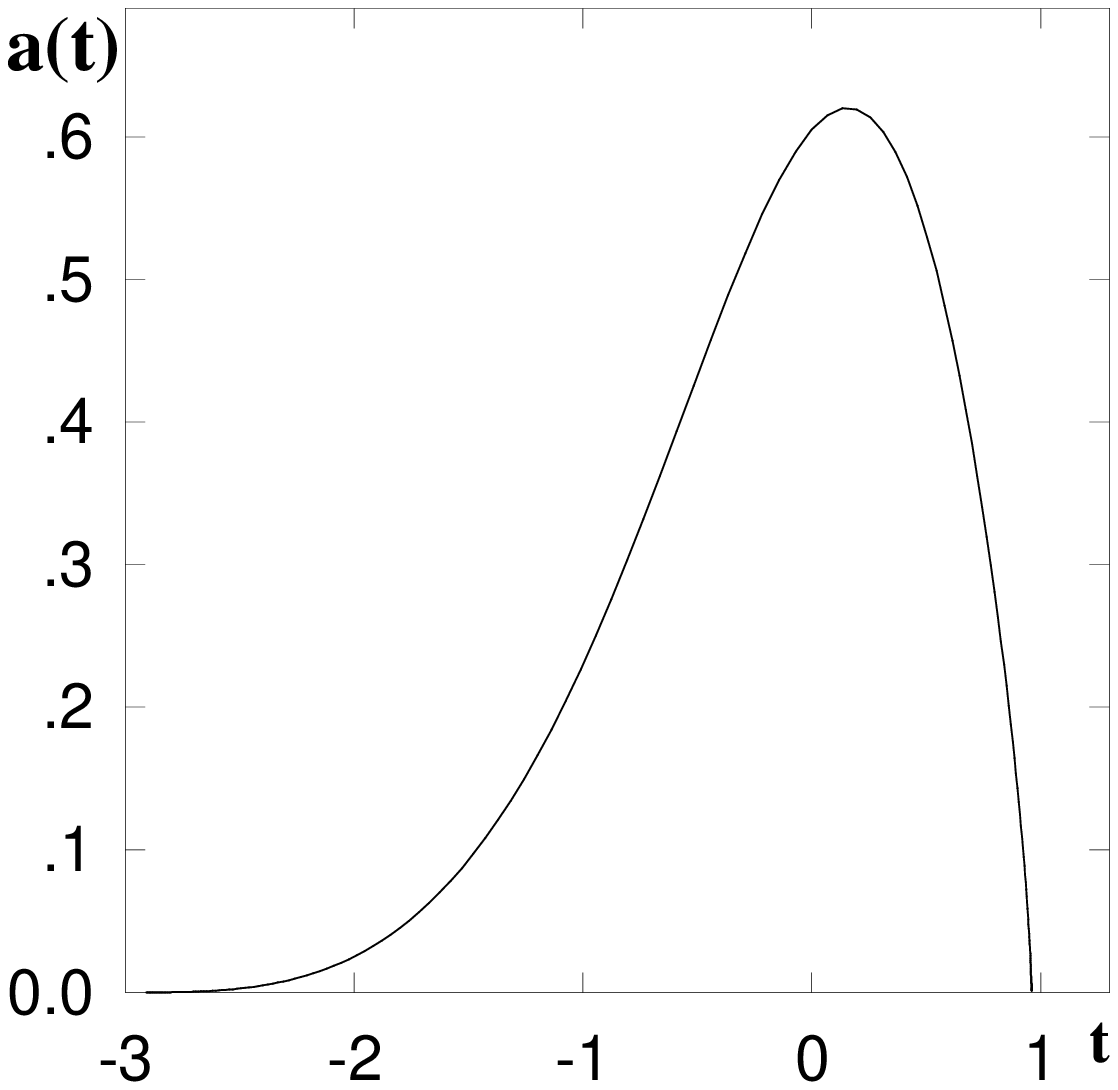,width=5cm,height=8cm}}
      \put(-10,30){\epsfig{file=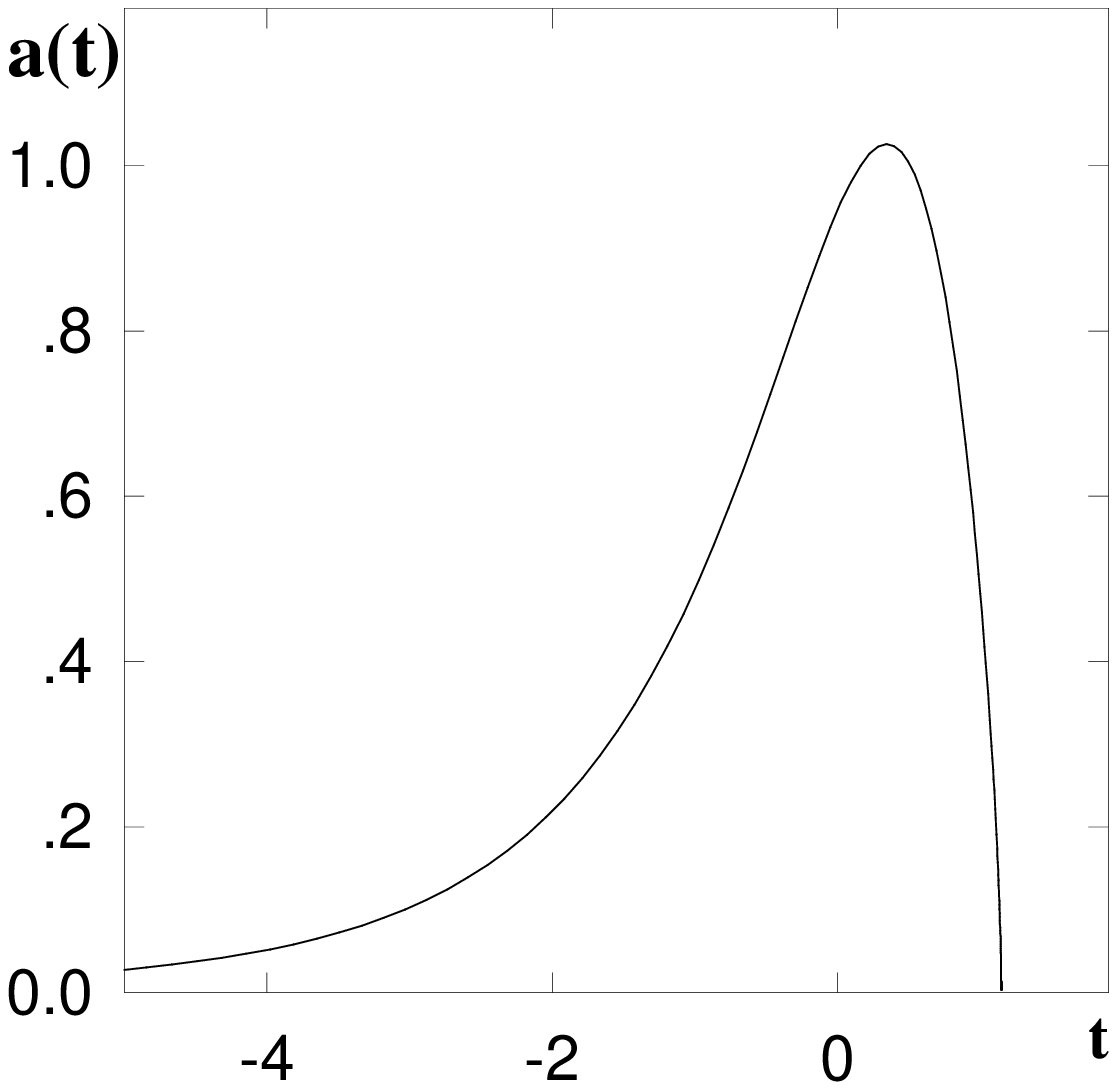,width=5cm,height=8cm}}
      \put(30,30){\epsfig{file=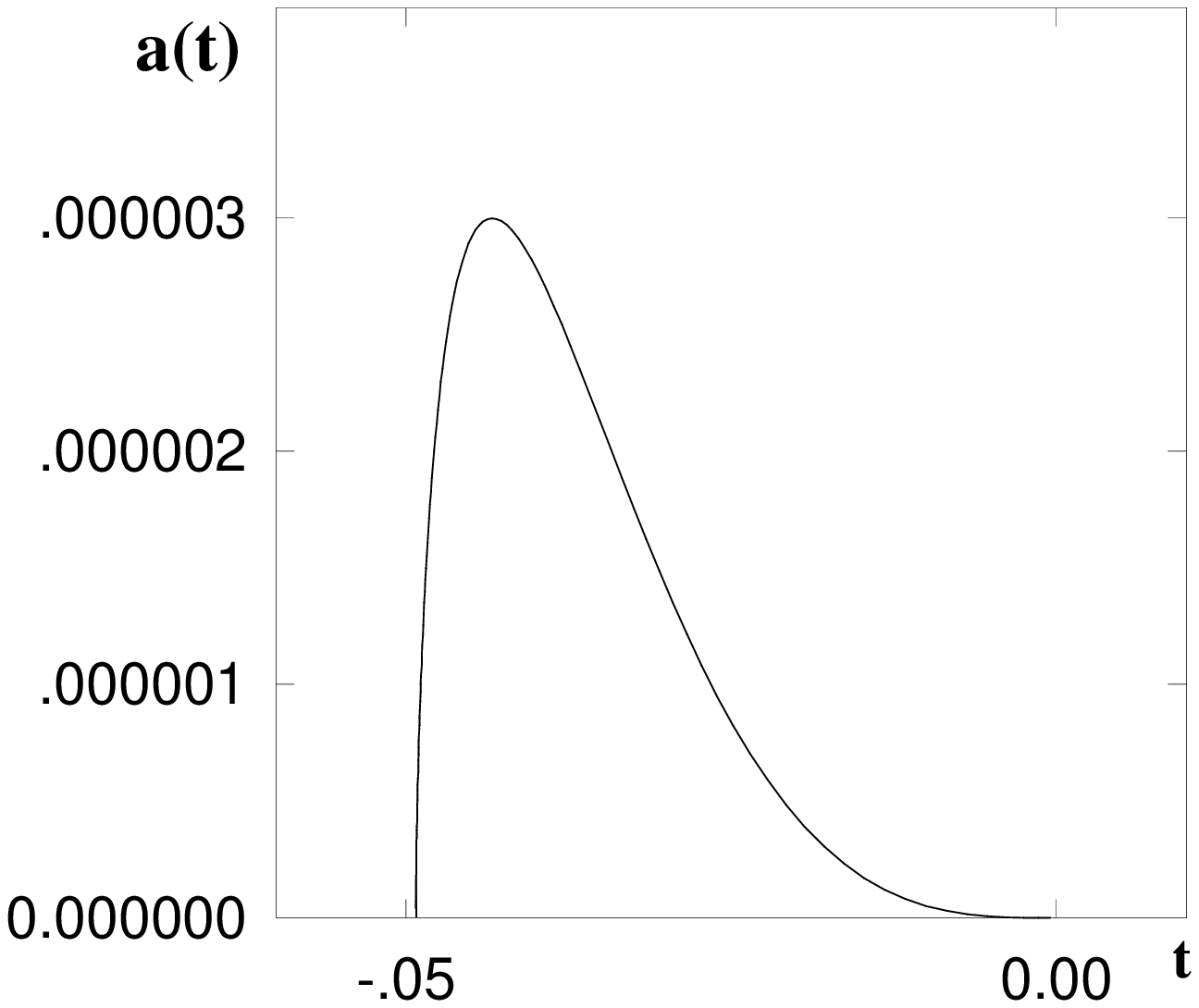,width=5cm,height=8cm}}
      \put(70,30){\epsfig{file=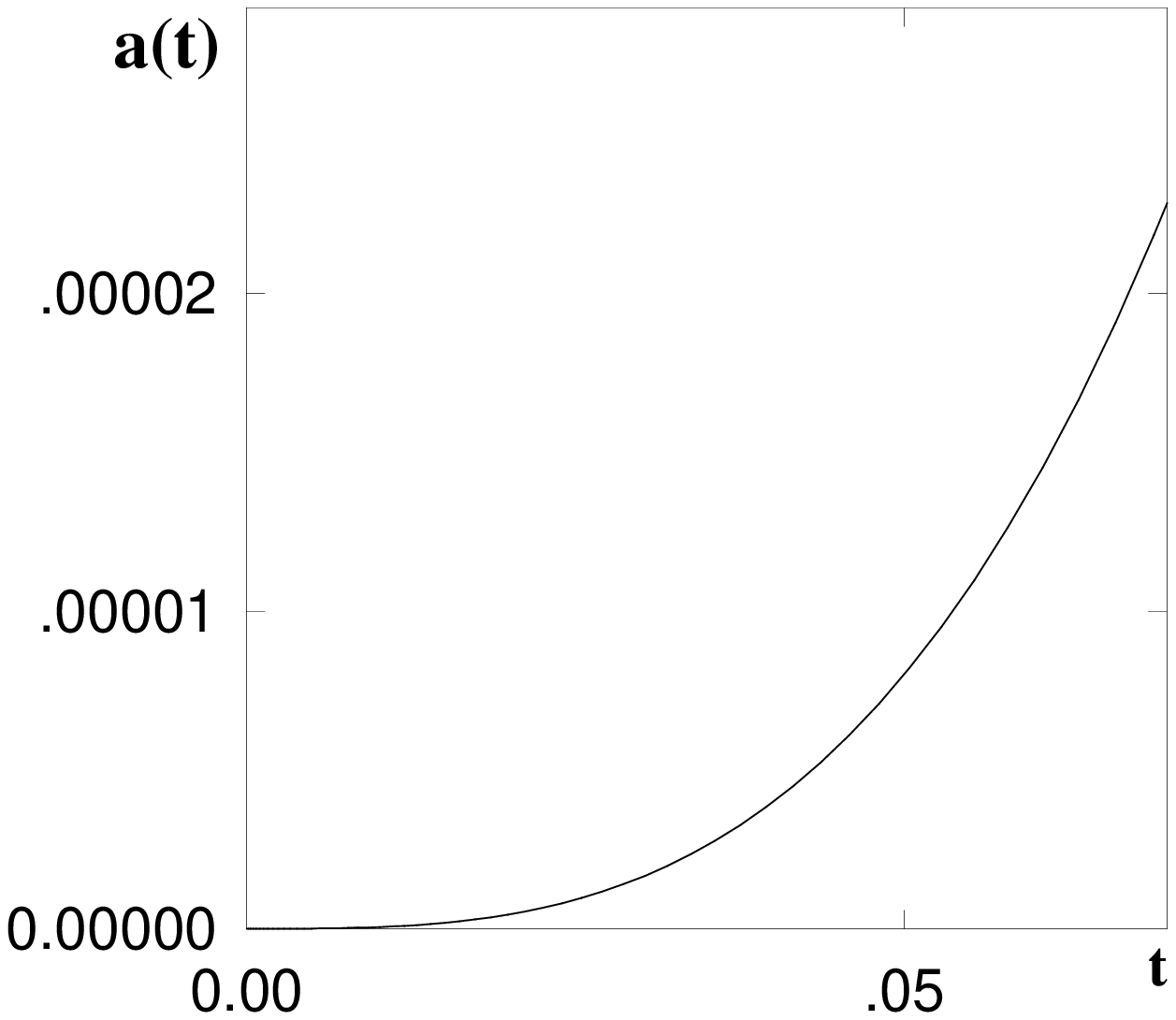,width=5cm,height=8cm}}
      \put(-50,-30){\epsfig{file=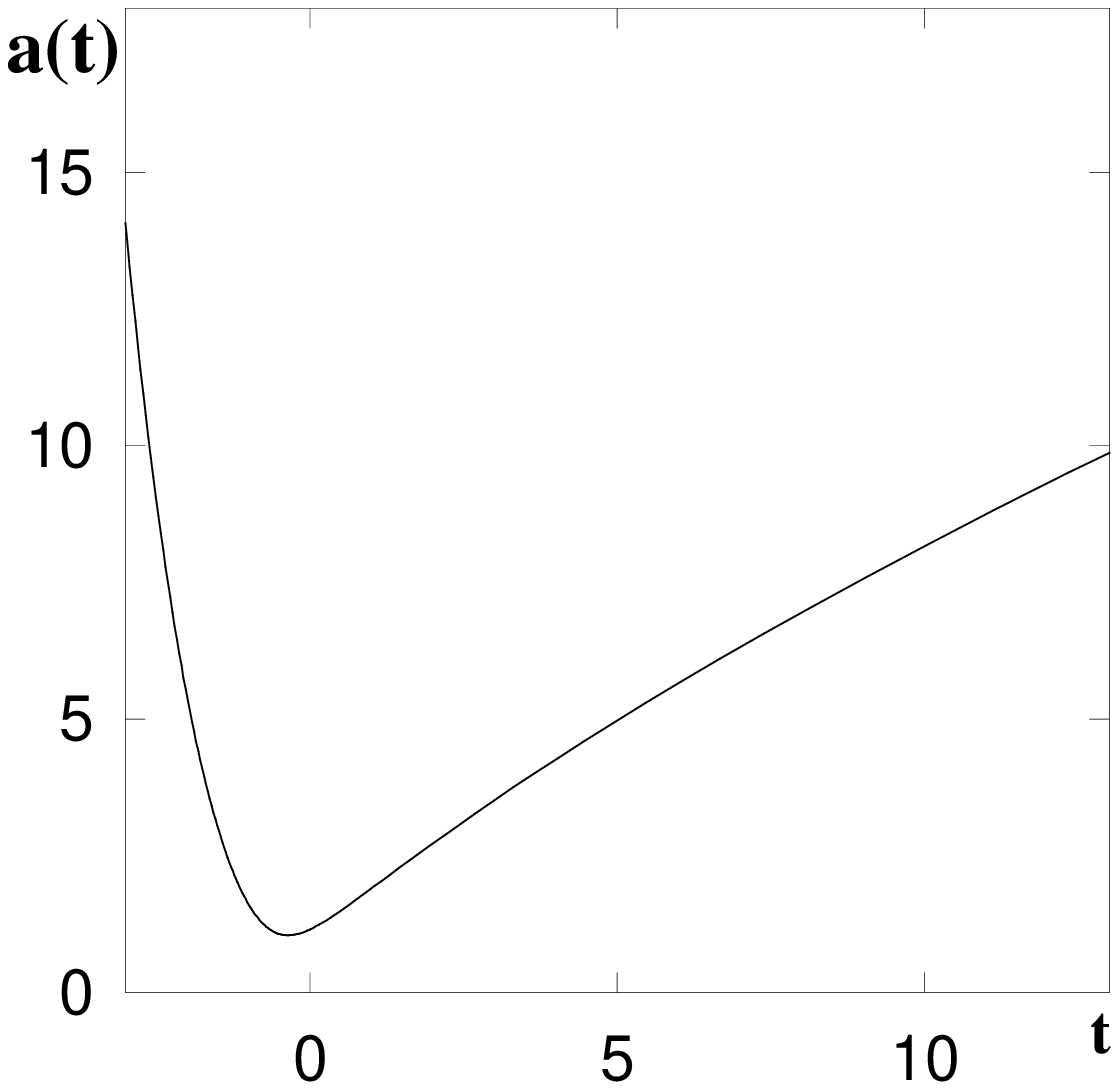,width=5cm,height=8cm}}
      \put(-10,-30){\epsfig{file=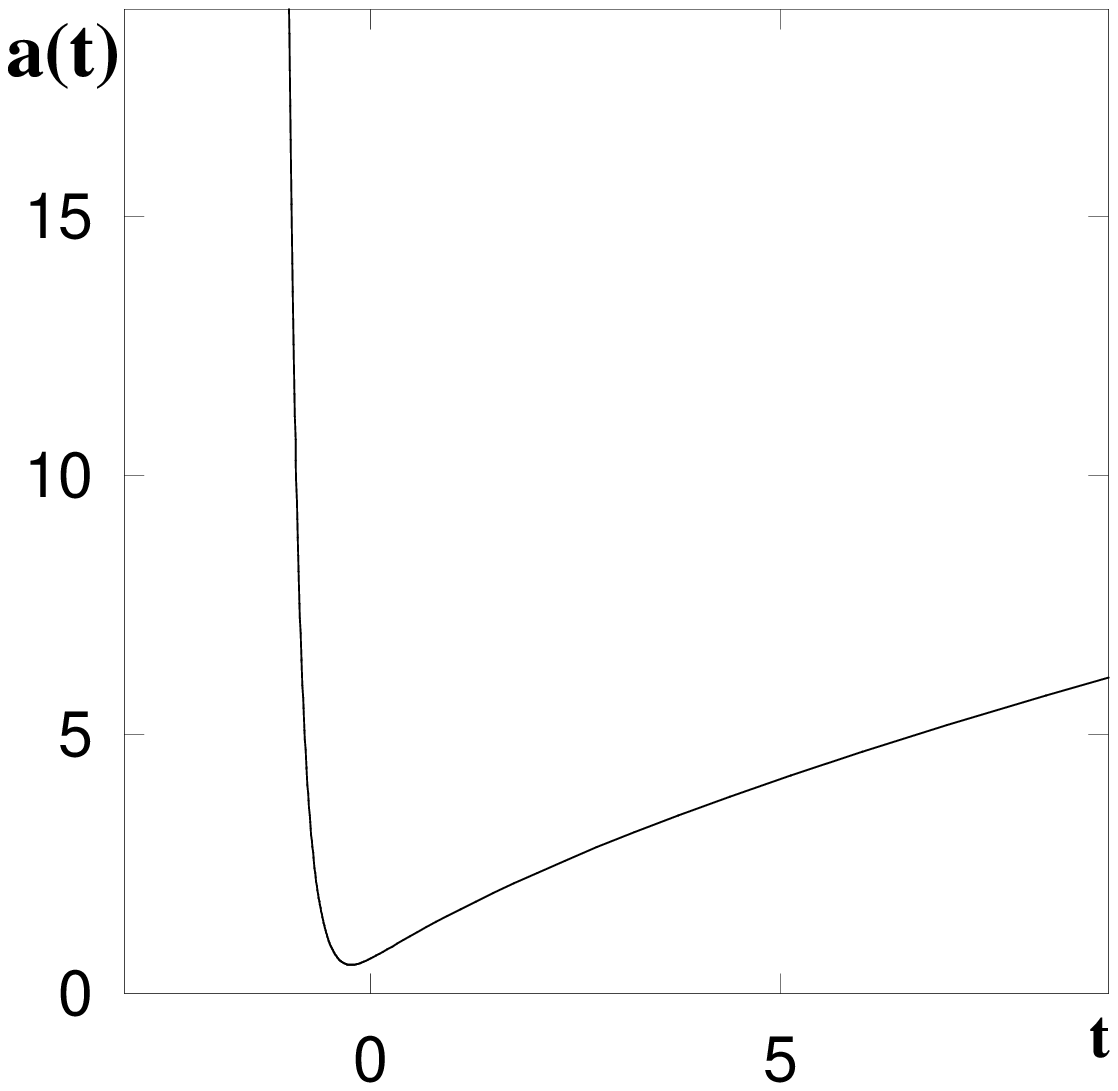,width=5cm,height=8cm}}
      \put(30,-30){\epsfig{file=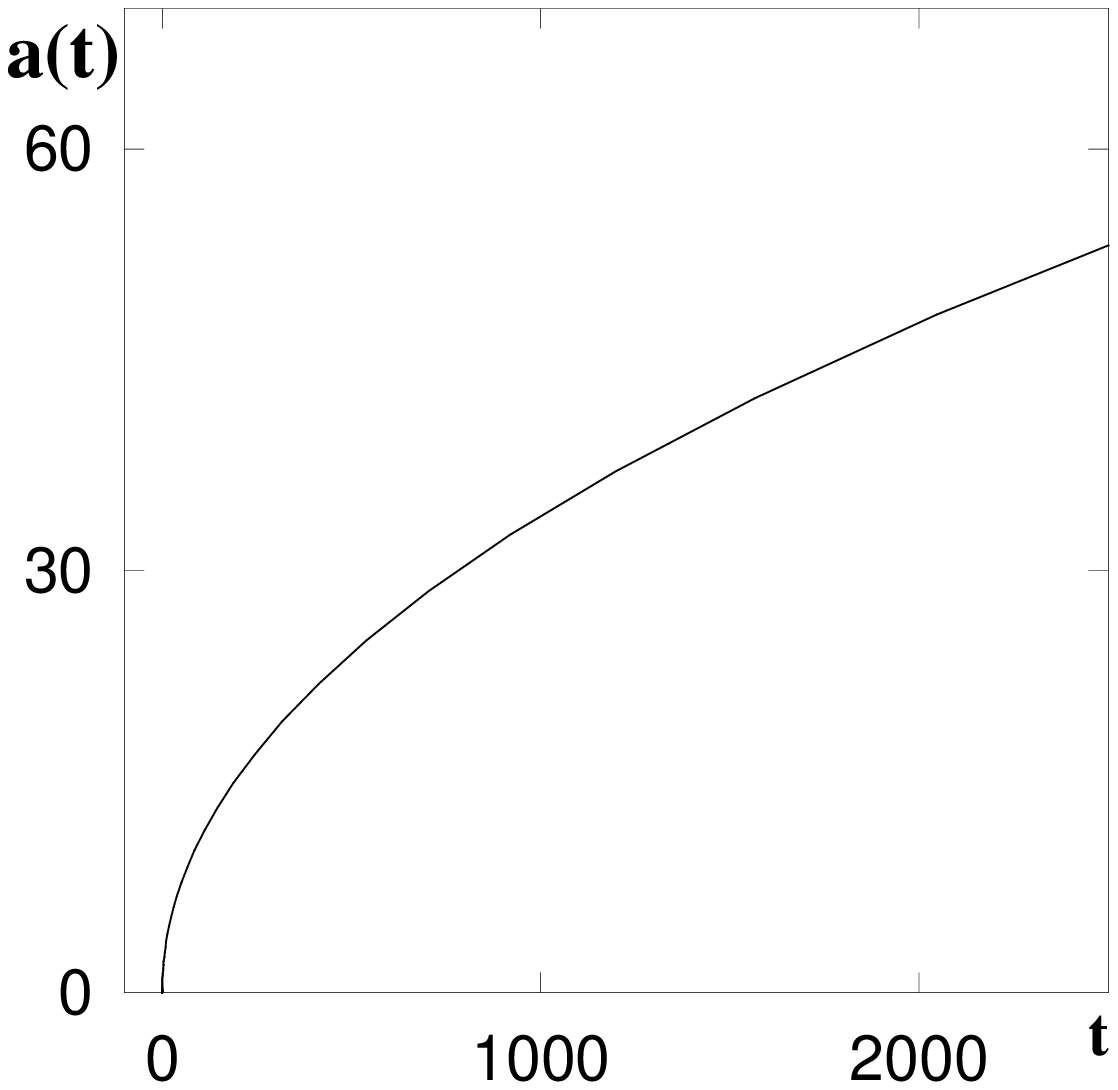,width=5cm,height=8cm}}
      \put(-35,40){$(a)$ phase $I$}
      \put(5,40){$(b)$ phase $II$}
      \put(45,40){$(c)$ phase $III^-$}
      \put(85,40){$(d)$ phase $III^+$}
      \put(-35,-15){$(e)$ phase $IV$}
      \put(5,-15){$(f)$ phase $V$}
      \put(45,-15){$(g)$ phase $VI$}
      \end{picture}
   \end{center}
\vspace{2cm}
\caption{
The behavior of the scale from phase $I$ to $VI$
}

\label{fig2}
\end{figure}
\newpage
\vspace{-3cm}
\begin{figure}
  \unitlength 1mm
   \begin{center}
      \begin{picture}(70,100)
      \put(-50,30){\epsfig{file=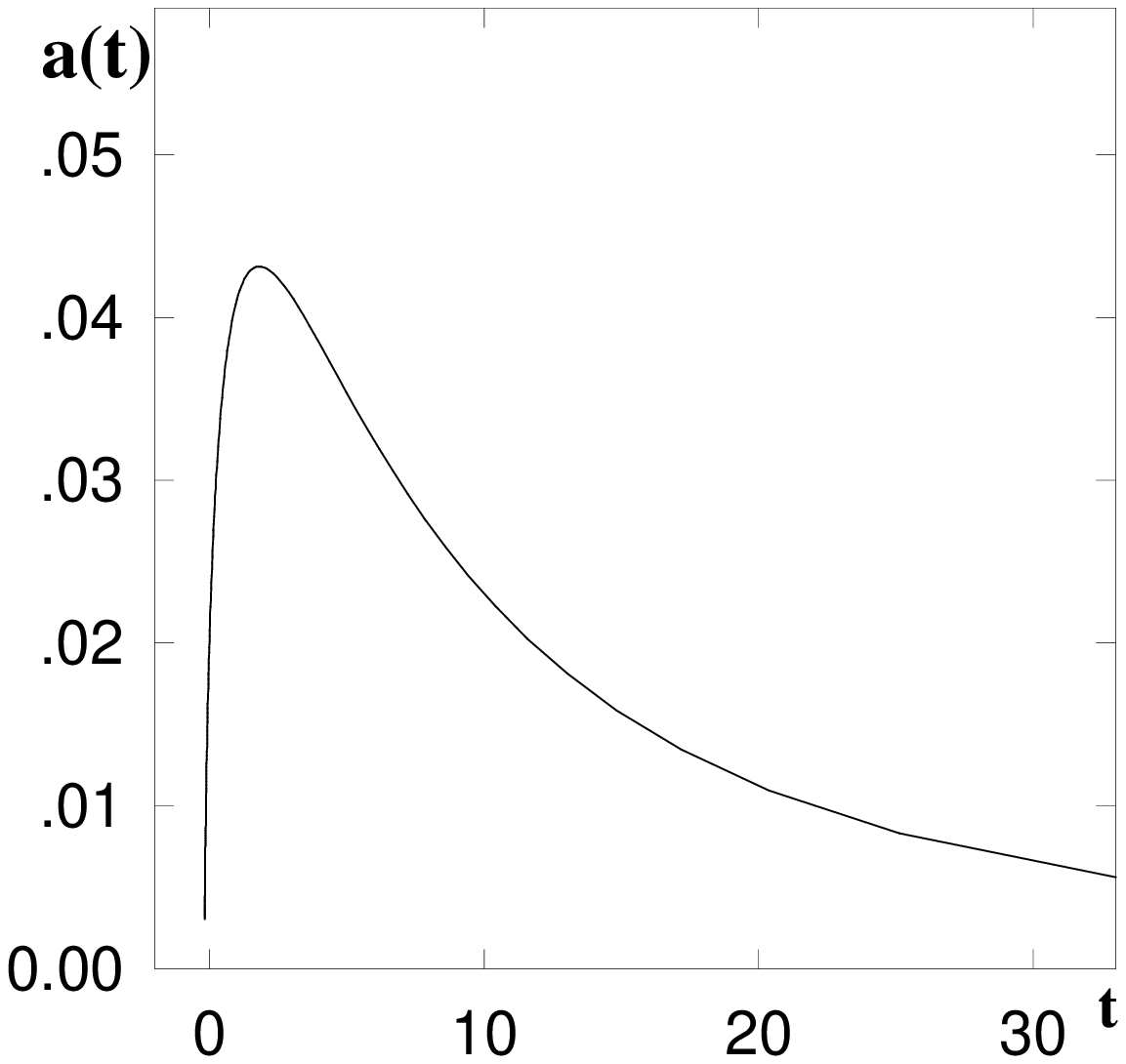,width=5cm,height=8cm}}
      \put(-10,30){\epsfig{file=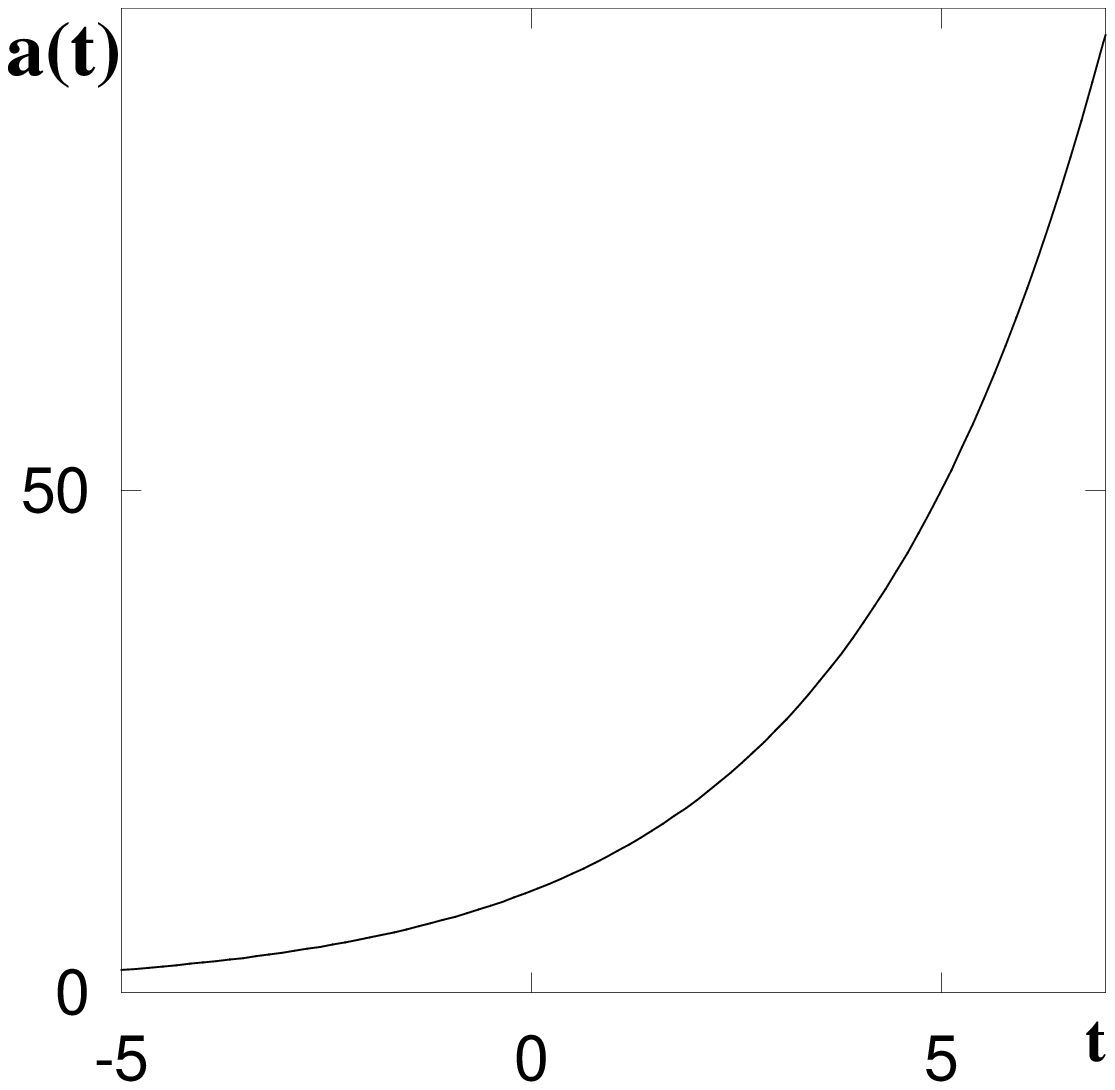,width=5cm,height=8cm}}

      \put(30,30){\epsfig{file=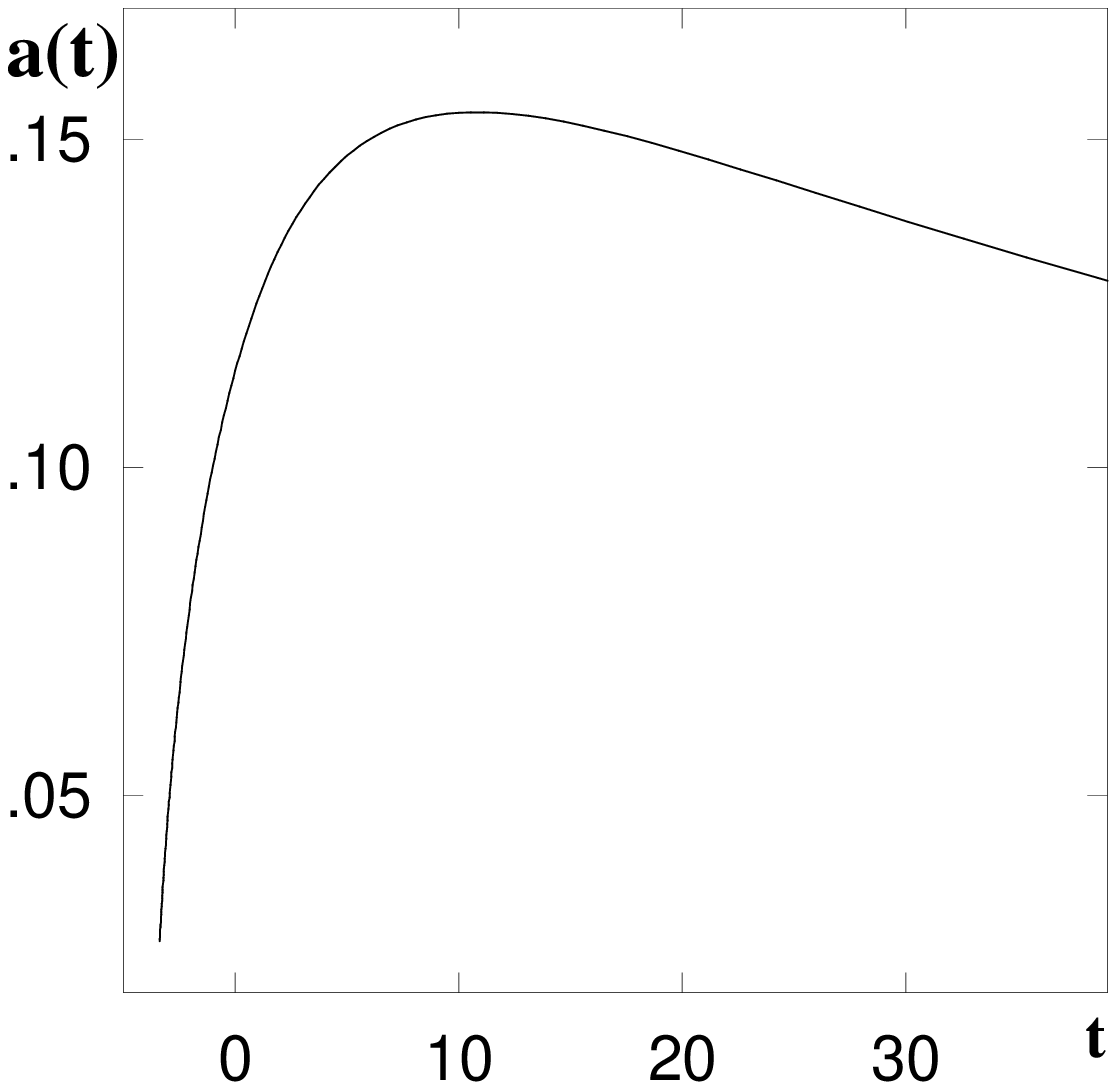,width=5cm,height=8cm}}
      \put(70,30){\epsfig{file=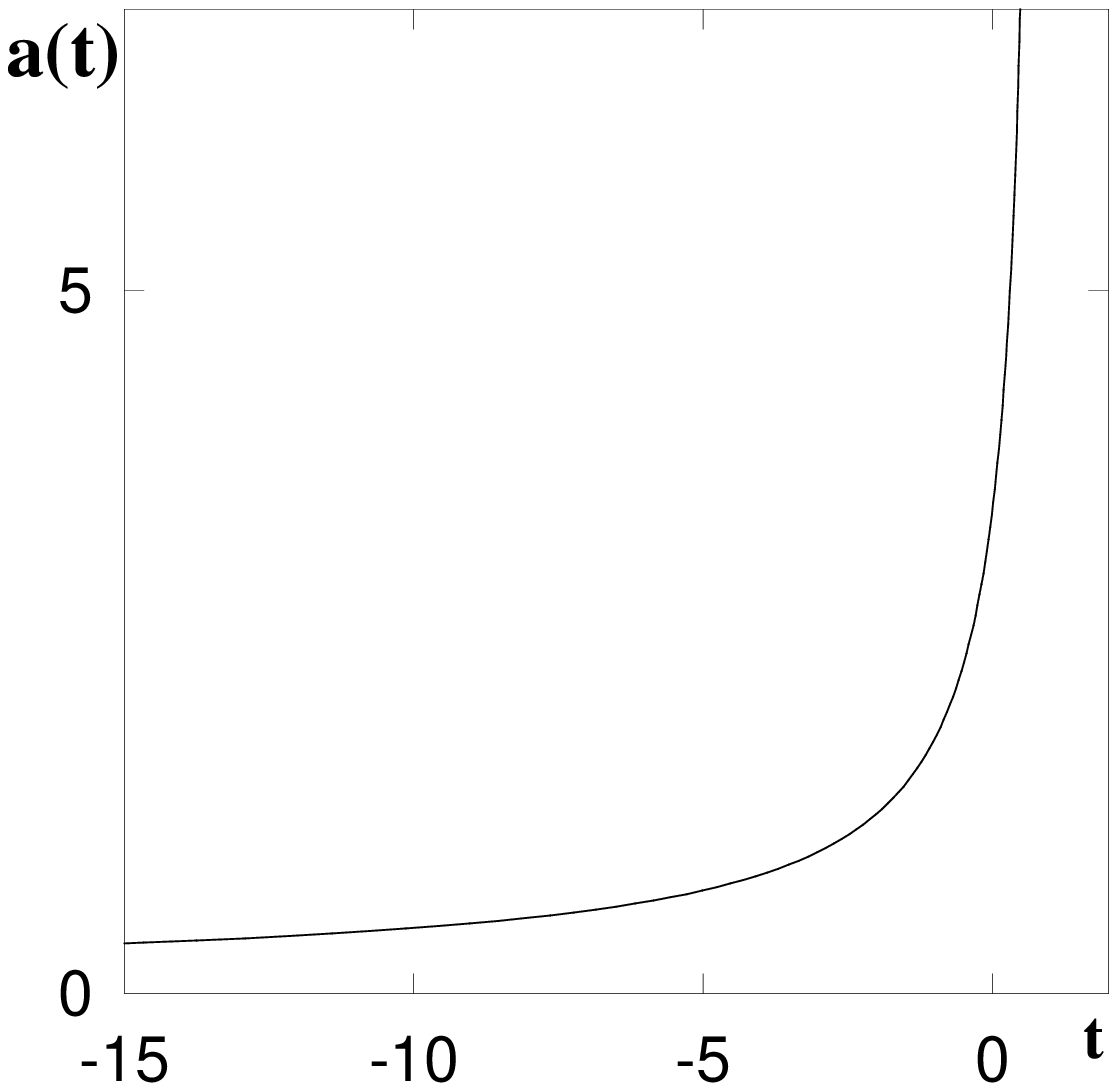,width=5cm,height=8cm}}
      \put(-50,-30){\epsfig{file=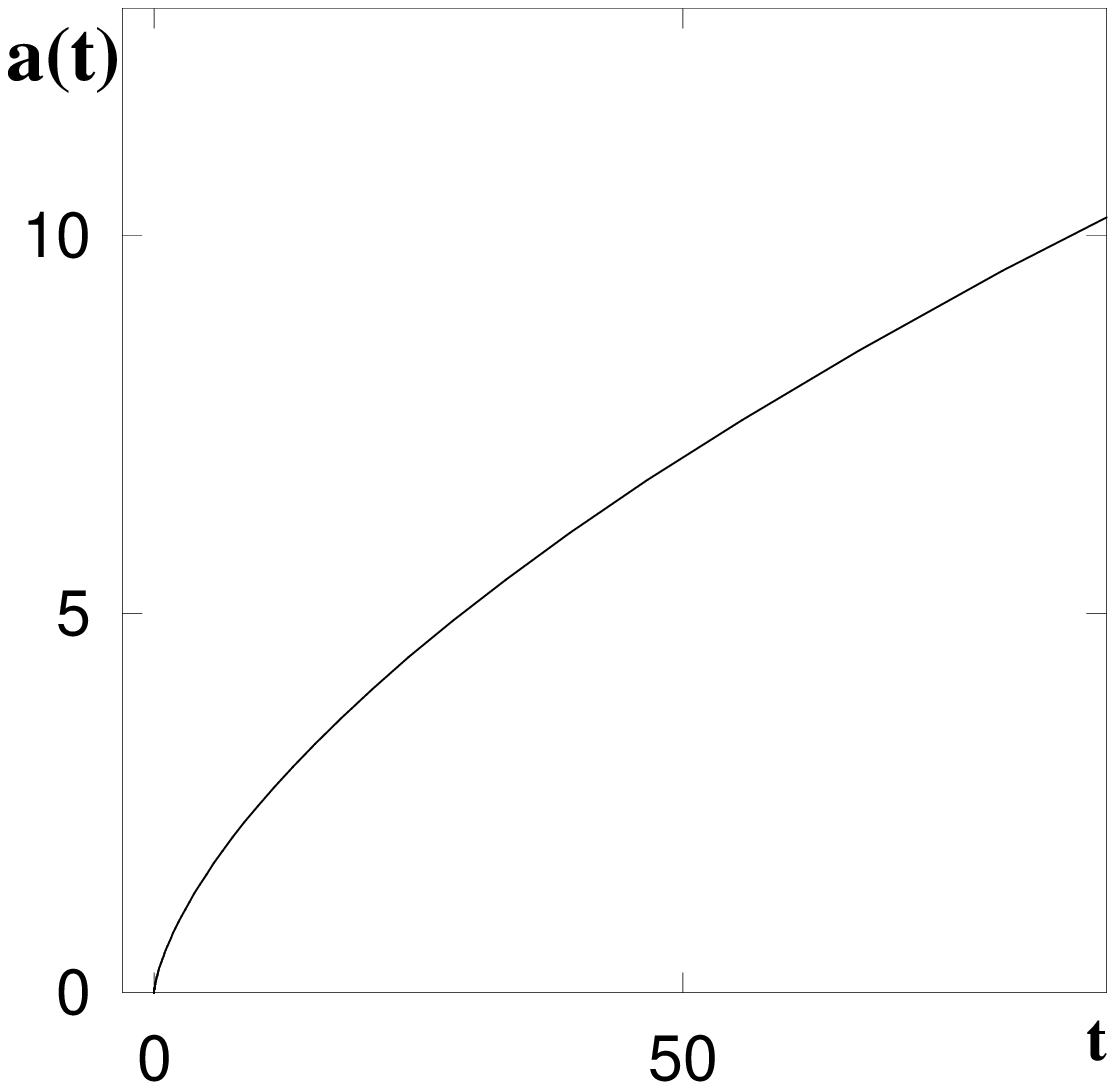,width=5cm,height=8cm}}
      \put(-10,-30){\epsfig{file=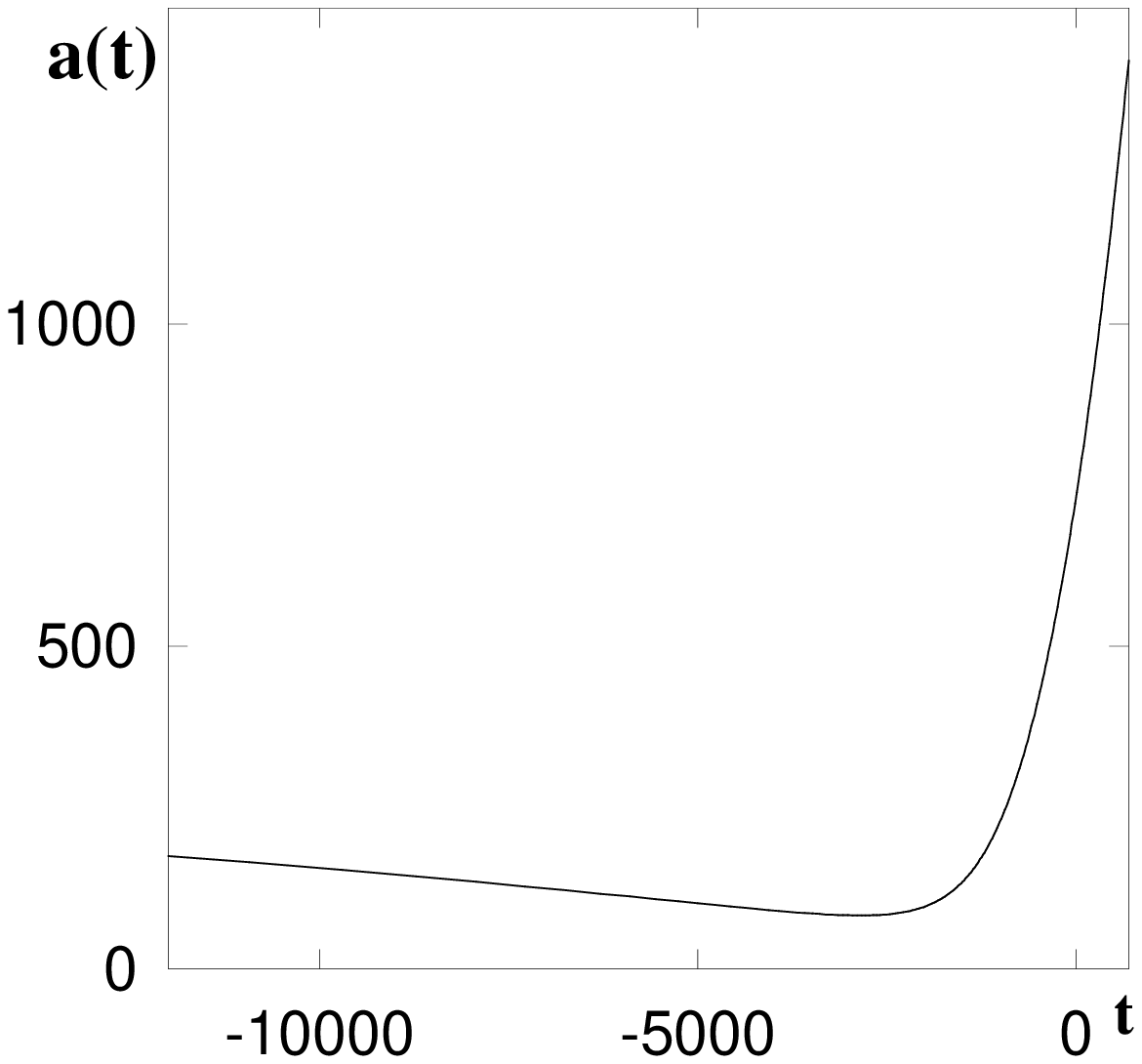,width=5cm,height=8cm}}
      \put(30,-30){\epsfig{file=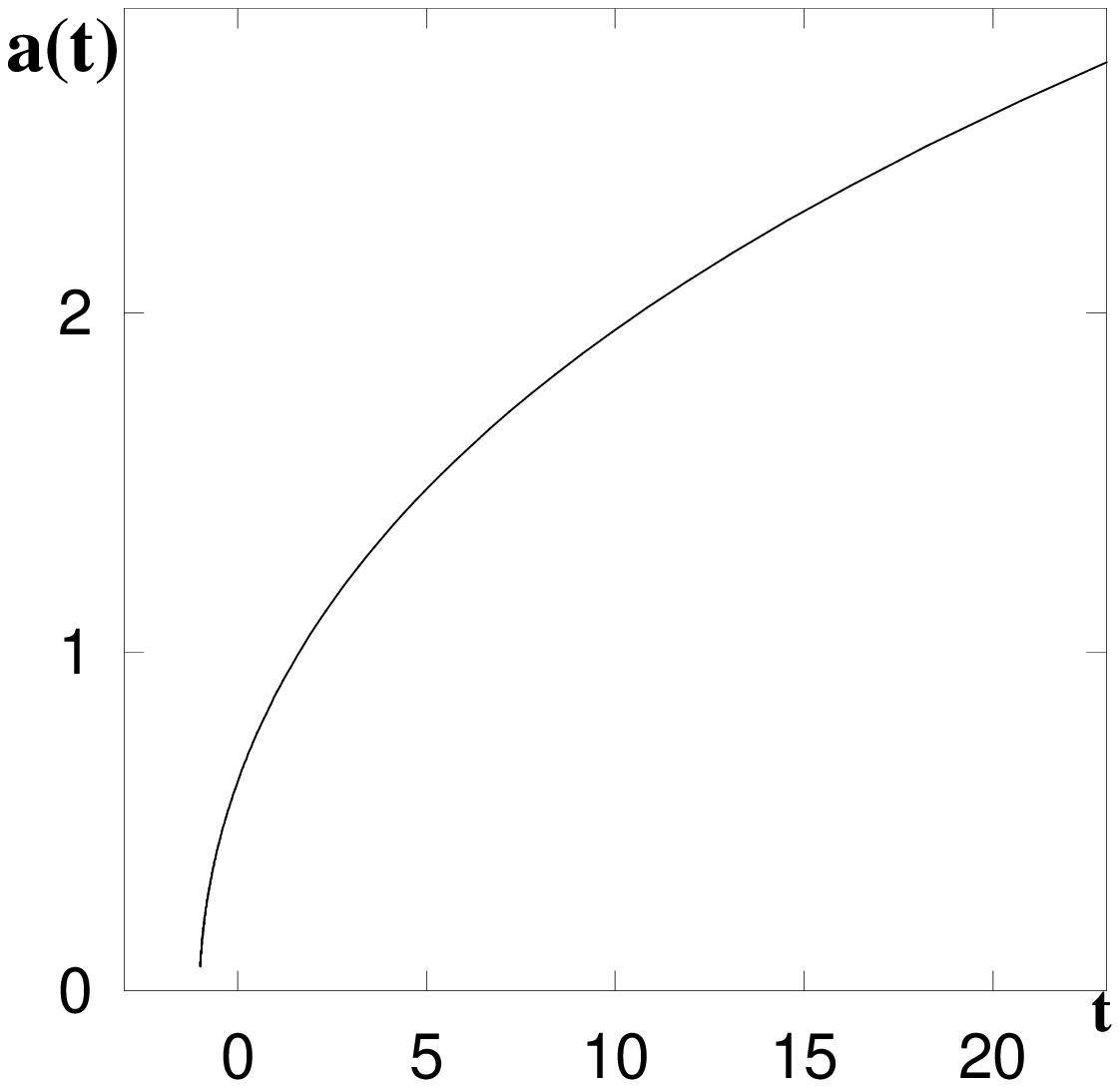,width=5cm,height=8cm}}
      \put(70,-30){\epsfig{file=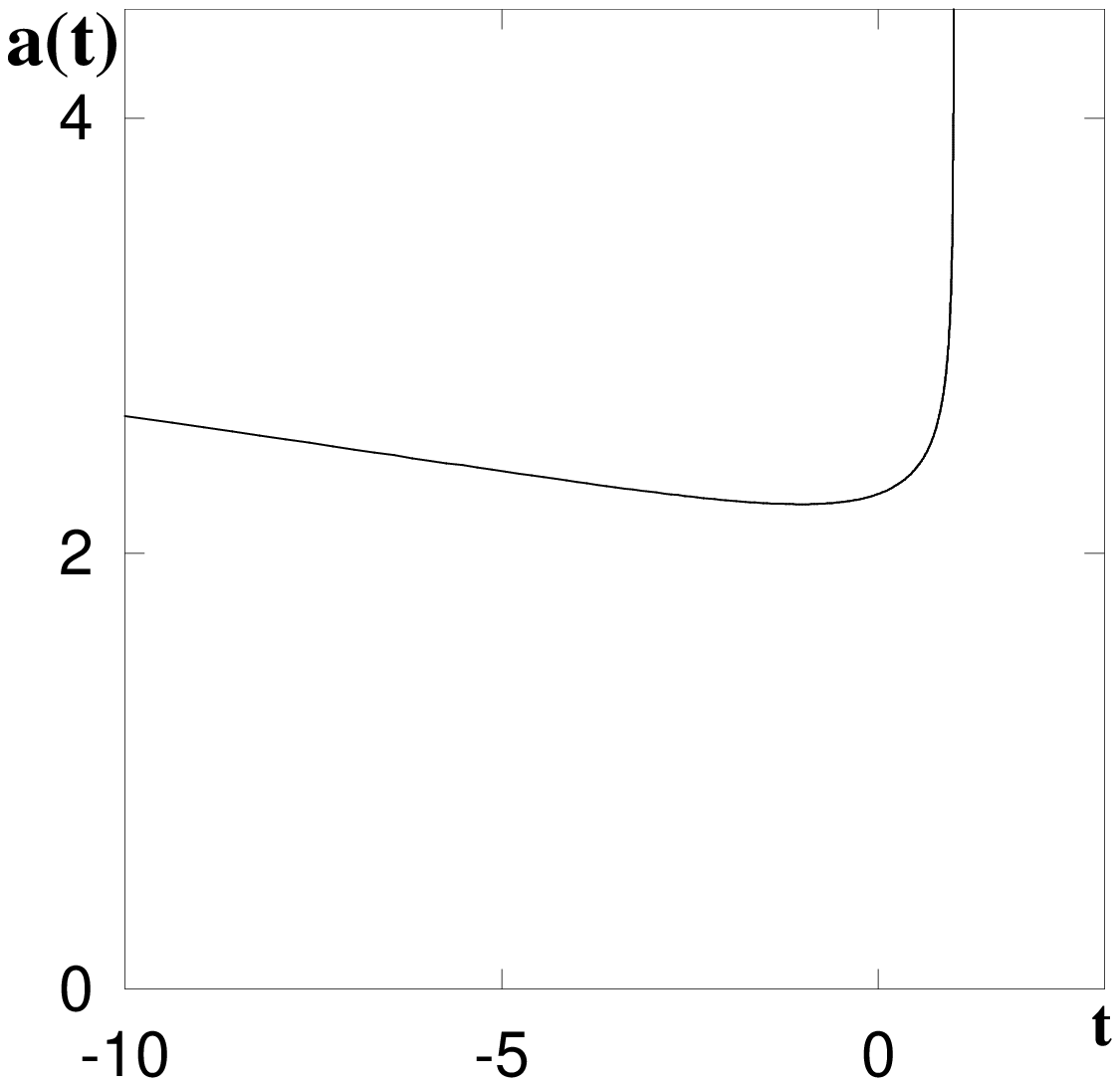,width=5cm,height=8cm}}
      \put(-35,40){$(a)$ phase $VII^-$}
      \put(5,40){$(b)$ phase $VII^+$}
      \put(45,40){$(c)$ phase $VIII^-$}
      \put(85,40){$(d)$ phase $VIII^+$}
      \put(-35,-20){$(e)$ phase $IX^-$}
      \put(5,-20){$(f)$ phase $IX^+$}
      \put(45,-20){$(g)$ phase $X^-$}
      \put(85,-20){$(h)$ phase $X^+$}
      \end{picture}
   \end{center}
\vspace{2.5cm}
\caption{
The behavior of scale factor from phase $VII^-$ to $X^+$
}
\label{fig3}
\end{figure}

\newpage
\vspace{-3cm}
\begin{figure}
  \unitlength 1mm
   \begin{center}
      \begin{picture}(70,100)
      \put(-50,30){\epsfig{file=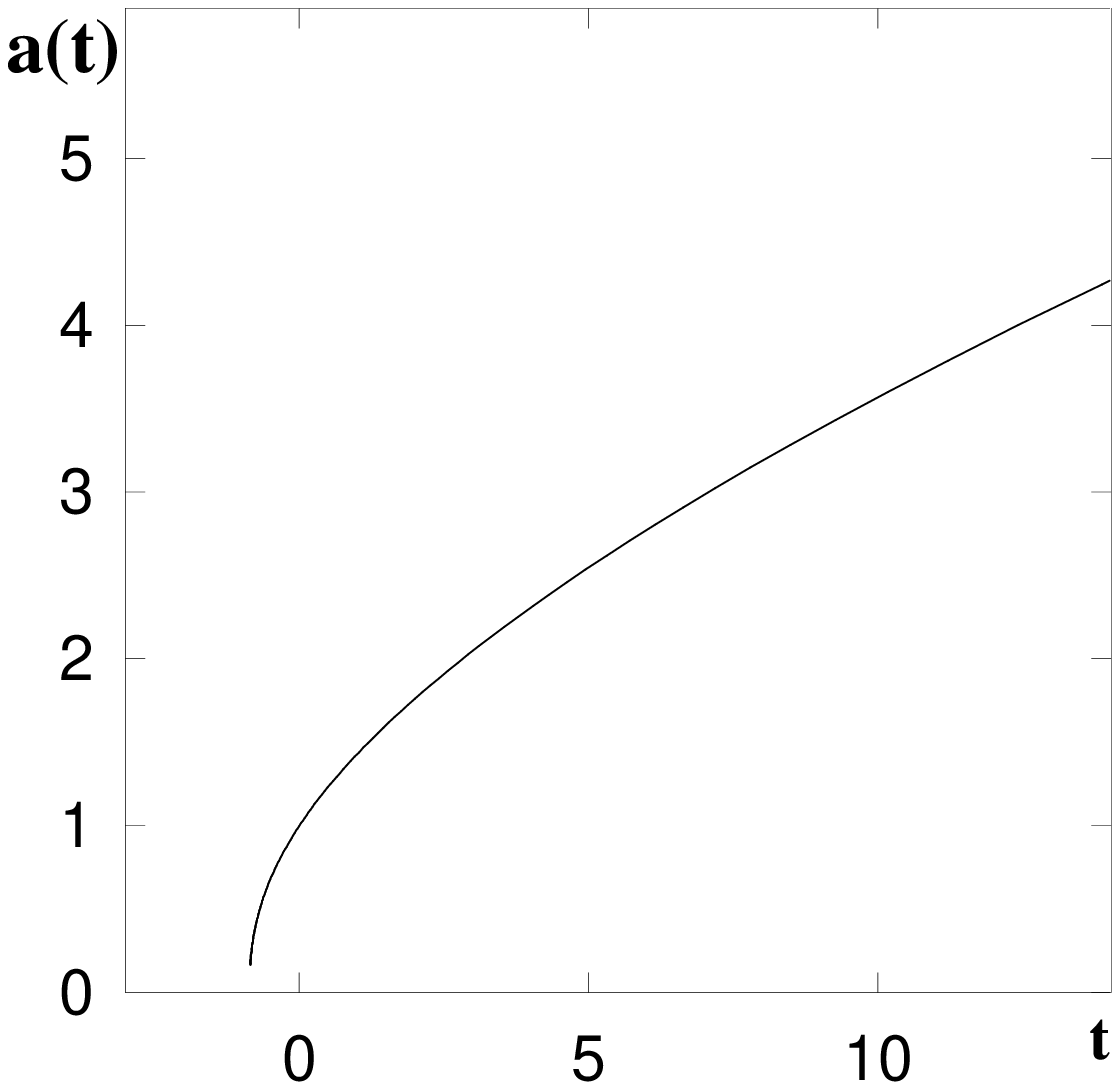,width=5cm,height=8cm}}
      \put(-10,30){\epsfig{file=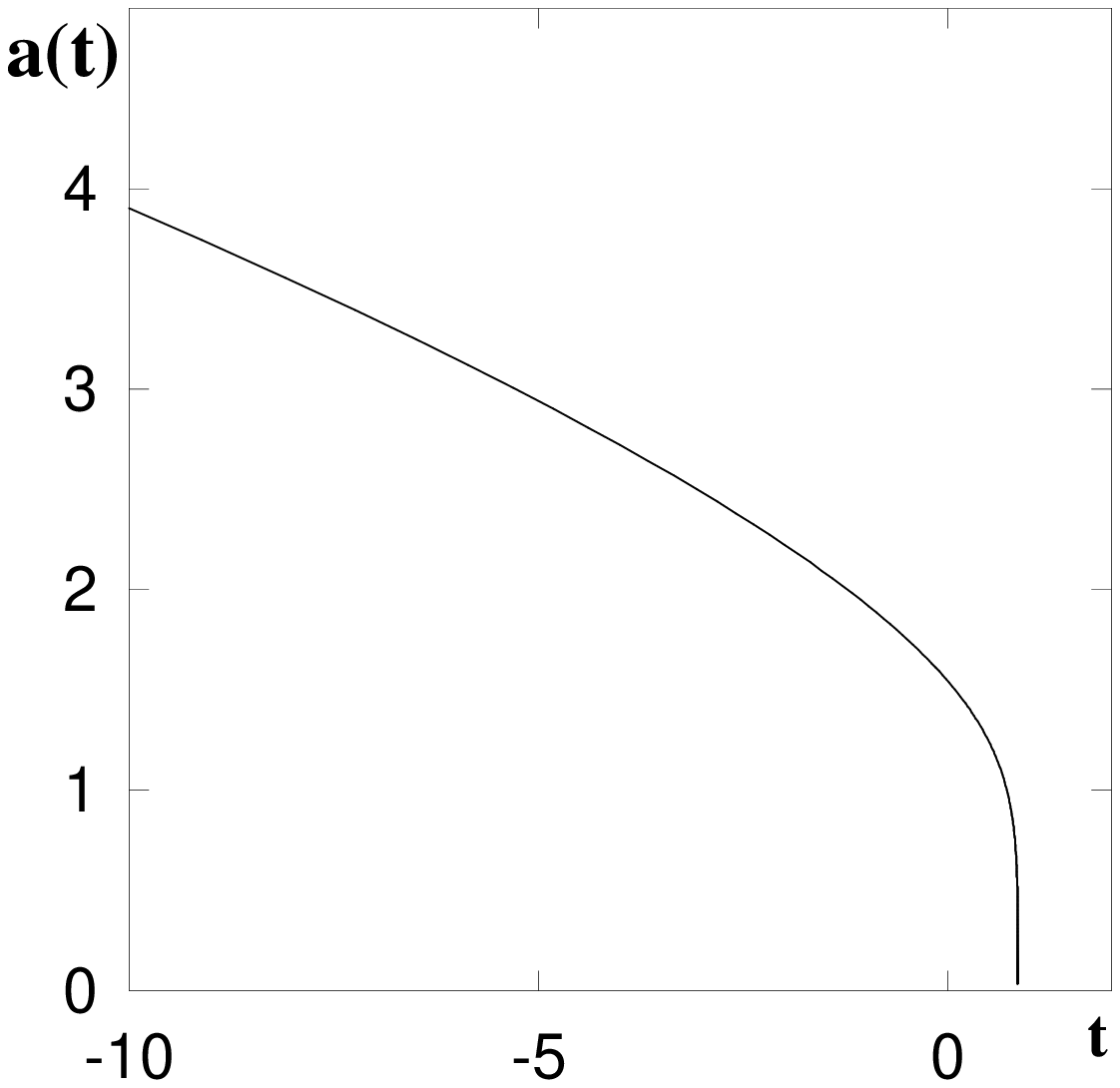,width=5cm,height=8cm}}
      \put(30,30){\epsfig{file=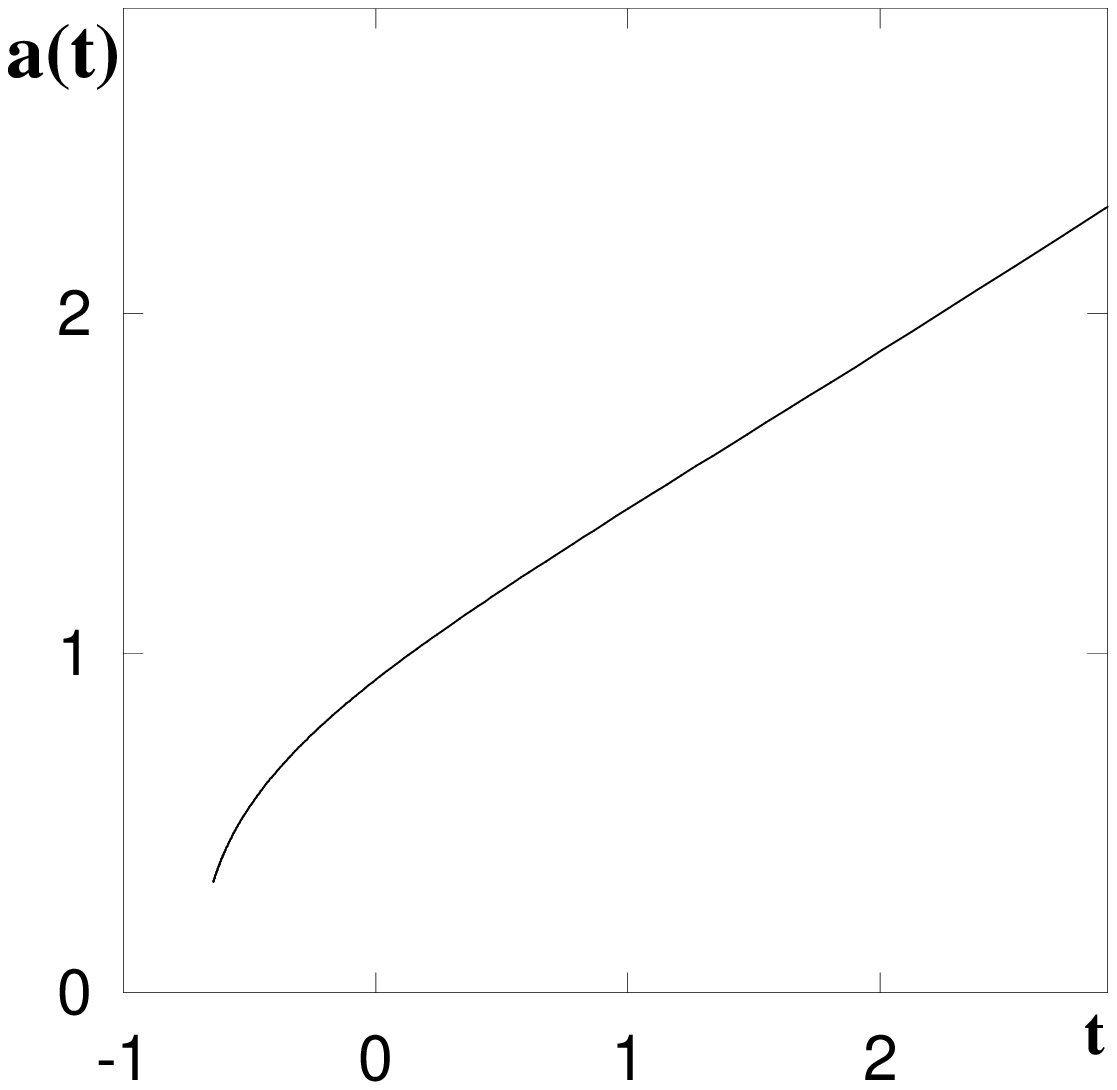,width=5cm,height=8cm}}
      \put(70,30){\epsfig{file=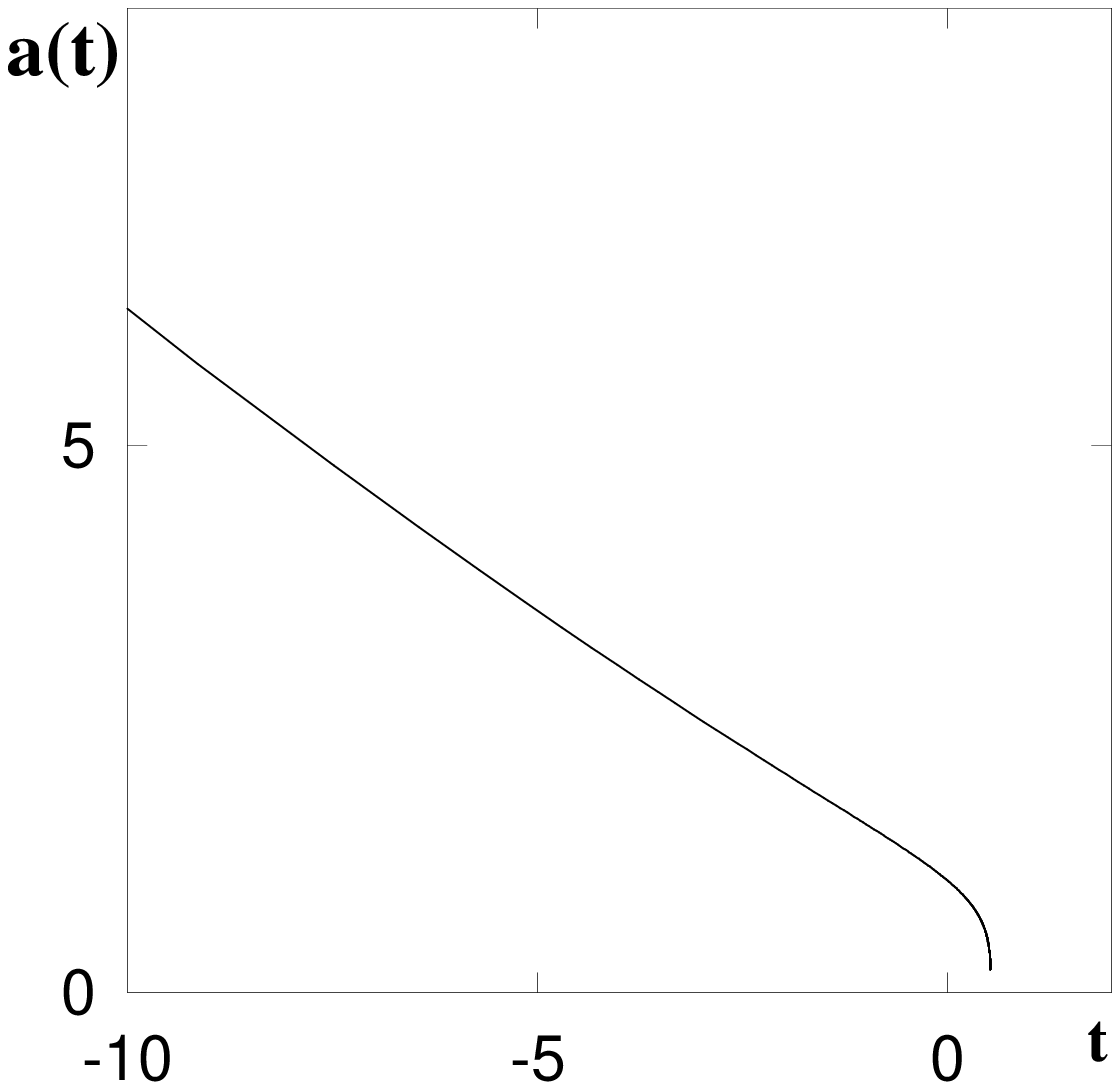,width=5cm,height=8cm}}

      \put(-35,40){$(a)$ phase $XI^-$}
      \put(5,40){$(b)$ phase $XI^+$}
      \put(45,40){$(c)$ phase $XII^-$}
      \put(85,40){$(d)$ phase $XII^+$}
      \end{picture}
   \end{center}
\vspace{-3.5cm}
\caption{
The behavior of scale factor from phase $XI^-$ to $XII^+$
}
\label{fig3}
\end{figure}

\vfil\eject
\end{document}